\documentstyle [12pt,amsfonts] {article}
\input epsf
\newcommand{\beq}{\begin{equation}}
\newcommand{\eeq}{\end{equation}}
\newcommand{\beqs}{\begin{eqnarray}}
\newcommand{\eeqs}{\end{eqnarray}}
\newcommand{\lsim}{\mathrel{\raisebox{-.6ex}{$\stackrel{\textstyle<}{\sim}$}}}

\newcommand{\gtwid}{\mathrel{\raise.3ex\hbox{$>$\kern-.75em\lower1ex
\hbox{$\sim$}}}}
\newcommand{\ltwid}{\mathrel{\raise.3ex\hbox{$<$\kern-.75em\lower1ex
\hbox{$\sim$}}}}

\catcode`@=11
\@addtoreset{equation}{section}
\@addtoreset{equation}{subsection}
\def\theequation{\ifnum\value{section}=0 \arabic{equation}\ignorespaces
\else \ifnum\value{section}=-1 A.\arabic{equation}\ignorespaces
\else \ifnum\value{subsection}=0
\thesection.\arabic{equation}\ignorespaces
\else \thesection.\arabic{subsection}.\arabic{equation}\ignorespaces
                           \fi
                      \fi
                 \fi}
\catcode`@=12

\newtheorem{th}{Theorem}[section]
\newtheorem{cor}{Corollary}[section]

\begin{document}

\begin{titlepage}

\begin{center}

{\Large \bf Flow Polynomials and their Asymptotic Limits for Lattice Strip
Graphs} 

\vspace{1.1cm}
{\large Shu-Chiuan Chang${}$\footnote{email:
shu-chiuan.chang@sunysb.edu}
and Robert Shrock${}$\footnote{email:
robert.shrock@sunysb.edu.  Dedicated to F. Y. Wu on his 70'th Birthday}}\\
\vspace{18pt}
 C. N. Yang Institute for Theoretical Physics \\
 State University of New York \\
 Stony Brook, NY 11794-3840 \\
\end{center}

\vskip 0.6 cm

\begin{abstract}
\vspace{2cm}

We present exact calculations of flow polynomials $F(G,q)$ for lattice strips
of various fixed widths $L_y$ and arbitrarily great lengths $L_x$, with several
different boundary conditions.  Square, honeycomb, and triangular lattice
strips are considered. We introduce the notion of flows per face $fl$ in the
infinite-length limit.  We study the zeros of $F(G,q)$ in the complex $q$ plane
and determine exactly the asymptotic accumulation sets of these zeros ${\cal
B}$ in the infinite-length limit for the various families of strips.  The
function $fl$ is nonanalytic on this locus.  The loci are found to be
noncompact for many strip graphs with periodic (or twisted periodic)
longitudinal boundary conditions, and compact for strips with free longitudinal
boundary conditions.  We also find the interesting feature that, aside from the
trivial case $L_y=1$, the maximal point, $q_{cf}$, where ${\cal B}$ crosses the
real axis, is universal on cyclic and M\"obius strips of the square lattice for
all widths for which we have calculated it and is equal to the asymptotic value
$q_{cf}=3$ for the infinite square lattice.

\end{abstract}
\end{titlepage}

\section{Introduction}

Consider a connected graph $G=(V,E)$ with vertex set $V$ and edge (bond) set
$E$ and an abelian group $H$ of order $o(H)$, represented as an additive group.
For some general results, it will be necessary to allow the possibility of
multiple edges joining a given pair of vertices, and loops (edges joining
vertices to themselves), but these will usually not be present on the strip
graphs of interest here. For definiteness, we let $H$ be the additive group of
integers mod $q$, ${\mathbb Z}_q$.  Denote the number of vertices and edges as
$n=|V|$ and $|E|$.  Assign an orientation to each of the edges in $G$ and
consider a mapping that assigns to each of these oriented edges a nonzero
element in ${\mathbb Z}_q$.  A flow on $G$ is then defined as an assignment of
this type that satisfies the condition that the flow into each vertex is equal
to the flow outward from this vertex.  Here the addition of two flows is
defined according to the abelian group $H$, so that, for $H={\mathbb Z}_q$, the
flow into each vertex is equal to the flow out of this vertex mod ${\mathbb
Z}_q$.  In a fluid or electric circuit analogy, this means that the
(discretized) flow or electric current is conserved at each vertex mod $q$;
there are no sinks or sources.  Since the assignment of the zero element of
${\mathbb Z}_q$ to a given oriented edge is equivalent to the absence of the
edge as far as the flow is concerned, a standard restriction is that an
admissible flow must avoid zero flow numbers on any edge; this is termed a
nowhere-zero $q$-flow.  Henceforth, since all of our discussion will concern
nowhere-zero $q$-flows, we shall take ``$q$-flow'' to mean ``nowhere-zero
$q$-flow''.  Some works discussing flow polynomials include
\cite{tutte50}-\cite{rwt}.

An important problem in mathematical graph theory concerns the enumeration of
$q$-flows on a given (connected) graph $G$.  Tutte showed that there exists a
polynomial in $q$, which we denote $F(G,q)$, that equals this number of
$q$-flows \cite{tutte54,tutte84}.  The existence of this polynomial relies upon
the fact that, for a given abelian group $H$, the number of flows on $G$
depends only on the order of $H$, $o(H)=q$, not on any other structural
properties of $H$.  Clearly if $G$ contains a bridge (isthmus), then there are
no $q$-flows since the fluid or electric current flowing across the bridge has
no way of returning.  Thus, for any such edge that is a bridge, denoted $e_b$,
one has $F(e_b,q)=0$.  The minimum (integer) value of $q$ such that a
bridgeless graph $G$ admits a $q$-flow is called the flow number of $G$,
denoted $\phi(G)$.

In this paper we shall present exact calculations of flow polynomials for a
number of lattice strips with various widths $L_y$ and arbitrarily great
lengths $L_x$m having several different types of boundary conditions.  The
lattices considered include the square ($sq$), honeycomb ($hc$), and triangular
($tri$) lattices.  The longitudinal (transverse) direction is taken as the
horizontal, $x$ (vertical, $y$) direction.  We envision the strip of the
triangular lattice as being formed by starting with a strip of the square
lattice and then adding edges joining, say, the lower left and upper right
vertices of each square.  The strips of the honeycomb lattice are envisioned as
brick lattices.  We use the symbols FBC$_y$ and PBC$_y$ for free and periodic
transverse boundary conditions and FBC$_x$, PBC$_x$, and TPBC$_x$ for free,
periodic, and twisted periodic longitudinal boundary conditions.  The term
``twisted'' means that the longitudinal ends of the strip are identified with
reversed orientation. These strip graphs can be embedded on surfaces with the
following topologies: (i) (FBC$_y$,FBC$_x$): free or open strip; (ii)
(PBC$_y$,FBC$_x$): cylindrical; (iii) (FBC$_y$,PBC$_x$): cyclic; (iv)
(FBC$_y$,TPBC$_x$): M\"obius; (v) (PBC$_y$,PBC$_x$): torus; and (vi)
(PBC$_y$,TPBC$_x$): Klein bottle.

We shall introduce the notion of flows per face, $fl$ in the limit $|V| \to
\infty$.  From the basic definition one can generalize $q$ from ${\mathbb Z}_+$
to ${\mathbb R}$ or, indeed, ${\mathbb C}$.  This generalization is necessary
when one calculates the zeros of $F(G,q)$ in the complex $q$ plane.  Using our
exact calculations of $F(G,q)$ for a variety of families of lattice strip
graphs, we shall determine exactly the continuous accumulation set of these
zeros of $F(G,q)$ as $|V| \to \infty$ for each family of graphs $G$.  This
formal limit of the family of graphs $G$ as $|V| \to \infty$ is denoted
$\{G\}$. We shall call this accumulation set ${\cal B}$ or, when necessary to
distinguish it from other accumulation sets, ${\cal B}_{fl}$.  This locus is of
interest because the function $fl$ is nonanalytic across ${\cal B}$.  We also
study the approach of the $fl$ functions for various infinite-length strips of
the square lattice, as functions of $q$, to numerical values for the infinite
square lattice.

There are several motivations for this work.  The flow polynomial is of
fundamental importance in graph theory.  As we shall discuss below, it is a
special case of the Tutte polynomial and encodes information on the
connectivity of a graph.  It is also of interest from the viewpoint of
statistical mechanics because it is a special case of the $q$-state Potts model
partition function.  To our knowledge, there has not been any study of the
function $fl$ that we introduce, and the associated regions of analyticity of
$fl$ separated by the locus ${\cal B}$ for strips of regular graphs.  In
particular, the points where ${\cal B}$ crosses or intersects the real $q$ axis
are somewhat analogous to phase transitions in statistical mechanics, in the
sense that the $fl$ function has different analytic forms on different sides of
${\cal B}$.  Associated with this, one finds qualitatively different flow
behavior, as described by $fl$, in the different regions separated by the locus
${\cal B}$.  While the flow polynomial of a planar graph is equivalent to the
chromatic polynomial of its dual graph, as we shall discuss further below, we
find that the locus ${\cal B} \equiv {\cal B}_{fl}$ for the infinite-length
limit of a strip graph of a regular lattice with periodic (or twisted periodic)
longitudinal boundary conditions is rather different from the corresponding
locus of zeros ${\cal B}_W$ of the chromatic polynomial for the infinite-length
limit of this strip.  These differences are especially intriguing in view of a
theorem that we shall give below, showing that in the limit of infinite width,
i.e. for the two-dimensional lattice $\Lambda$, ${\cal B}_{fl}(\Lambda)={\cal
B}_W(\Lambda^*)$, where $\Lambda^*$ is the planar lattice dual to $\Lambda$.

The flow polynomial also appears in applied mathematics, engineering, business,
and economics.  Important questions in engineering and business concern the
optimization of the flow capacity through parts of networks and the study of
where bottlenecks in flows can occur because of limited connectivity.  Although
we shall not have occasion to make use of it, we mention the ``max-flow,
min-cut'' theorem of Ford and Fulkerson \cite{ff}. In the present work, we
shall concentrate on intrinsic properties of flow polynomials and their
connections with statistical mechanics.

We comment on the relation with previous calculations of chromatic polynomials
(e.g., \cite{nagle}-\cite{sstri}).  Given the connection (\ref{fpdual}),
calculations of chromatic polynomials for families of planar lattice strip
graphs immediately yield flow polynomials for the dual graphs. We shall not
recapitulate these calculations here since they are available in the
literature.  Chromatic polynomials for nonplanar graphs cannot be immediately
related to flow polynomials.

\section{Connection with Tutte Polynomial and Potts Model}

In this section we review the connection between the flow polynomial and the
Tutte polynomial or equivalently the Potts model partition function.  Consider,
as before, a connected graph $G=(V,E)$.  If an edge of $G$ is a loop,
$e=e_\ell$, then the fluid (or electric current) conservation condition is
automatically satisfied for any value of $q$, so that $F(e_\ell,q)=q-1$, where
for (nowhere-zero) flows the zero element of $H$ is excluded and
there are thus $q-1$ choices for the flow through the loop.  We mention the
standard notation that for an edge $e \in E$, $G/e$ is the graph obtained from
$G$ by contraction on $e$, i.e., deleting the edge $e$ and identifying the
vertices that it joins, and $G-e$ is the graph obtained from $G$ by deleting
$e$.  Then for an edge $e$ that is not a bridge or a loop, the flow polynomial
satisfies the deletion-contraction relation
\beq
F(G,q) = F(G/e,q) - F(G-e,q) \ .
\label{fdelcon}
\eeq

If $G$ is a graph and $K$ is a field, then a general function $g: \
G \to K$ is a Tutte-Gr\"othendieck invariant if (1) if $e \in E$ is a bridge,
then $g(G)=g_b g(G/e)$, (2) if $e \in E$ is a loop, then
$g(G)=g_\ell g(G/e)$,
(3) if $e \in E$ is neither a bridge nor a loop, then
\beq
g(G) = a g(G-e)+b g(G/e) \quad a,b \ne 0 \ .
\label{dcrel}
\eeq
 From the discussion above, it follows that
$F(G,q)$ is a Tutte-Gr\"othendieck invariant.  A useful property is that a
Tutte-Gr\"othendieck invariant can be expressed in terms of the Tutte
polynomial.   A spanning subgraph $G^\prime=(V,E^\prime)$ of a graph $G=(V,E)$
is a subgraph with the same vertex set and a subset $E^\prime \subseteq E$ of
the edge set of $G$.  The Tutte polynomial of a graph $G$ is defined as 
\beq
T(G,x,y)=\sum_{G^\prime \subseteq G} (x-1)^{k(G^\prime)-k(G)}
(y-1)^{c(G^\prime)}
\label{t}
\eeq
where $k(G^\prime)$ and $c(G^\prime)$ denote the number of components and
linearly independent circuits of $G^\prime$, where, for an arbitrary graph 
$G=(V,E)$, 
\beq
c(G) = |E|-|V|+k(G)
\label{cg}
\eeq
The quantity $c(G^\prime)$ is also called the co-rank or nullity of $G^\prime$,
and the co-rank of the full graph $G$ is called its cyclomatic number.  The
rank of a graph $G$ is defined as $r(G)=|V|-k(G)$, so that the first exponent
can be written equivalently as $k(G^\prime)-k(G)=r(G)-r(G^\prime)$.  Since we
deal only with connected graphs here, $k(G)=1$.  (We follow the standard
notational usage of $x$ and $y$ for the arguments of the Tutte polynomial and
caution the reader not to confuse these with the $x$ and $y$ directions along
the strip graphs.)

A Tutte-Gr\"othendieck invariant $g$ is then given as
\beq
g(G) = a^{|E|-|V|+1}b^{|V|-1}T(G,\frac{g_b}{b},\frac{g_\ell}{a}) \
.
\label{tgtheorem}
\eeq
In particular, for the flow polynomial, in terms of the notation above, we have
$g_b=0$, $g_\ell = q-1$, $a=-1$, $b=1$, whence
\beq
F(G,q) = (-1)^{|E|-|V|+1}T(G,x=0,y=1-q) \ .
\label{ft}
\eeq
Thus, the flow polynomial is a special case of the Tutte polynomial. Hence, we
can use our previous exact calculations of Tutte polynomials and Potts model
partition functions for various families of graphs \cite{a}-\cite{ka3} to
derive flow polynomials for these graphs.

 From (\ref{t}) and (\ref{ft}), it follows that the degree of $F(G,q)$ as a
polynomial in $q$ is the cyclomatic number of $G$,
\beq
{\rm deg} \ F(G,q) = c(G) 
\label{degf}
\eeq
As is clear from the fact that the cyclomatic number $c(G)$ gives the number of
linearly independent circuits, it is closely related to the number of faces of
$G$, $f(G)$.  In particular, for a planar graph $G$, using the Euler relation
$|V|-|E|+f(G)=2$, we have $f(G)=c(G)+1$.

Next, we recall the equivalence of the Tutte polynomial to the Potts model
partition function.  On a lattice, or more generally, a graph $G$, at
temperature $T$, this model is defined by the partition function \cite{wurev}
\beq
Z(G,q,v) = \sum_{ \{ \sigma_n \} } e^{-\beta {\cal H}}
\label{zfun}
\eeq
with the Hamiltonian
\beq
{\cal H} = -J \sum_{\langle i j \rangle} \delta_{\sigma_i \sigma_j}
\label{ham}
\eeq
where $\sigma_i=1,...,q$ are the spin variables on each vertex $i \in G$, 
and $\langle i j \rangle$ denotes pairs of adjacent
vertices.  We use the notation
\beq
K = \frac{J}{k_BT} \ , \quad  v = e^K - 1
\label{kdef}
\eeq
(where there should not be any confusion between the variable $v$ and the edge
set $V$ or between the variable $K$ and the number of components of a graph
$k(G)$) so that the physical ranges are (i) $v \ge 0$ corresponding to $\infty
\ge T \ge 0$ for the Potts ferromagnet, and (ii) $-1 \le v \le 0$,
corresponding to $0 \le T \le \infty$ for the Potts antiferromagnet.

As before, let $G^\prime=(V,E^\prime)$ be a spanning subgraph of $G$. Then
$Z(G,q,v)$ can be written as \cite{fk}
\beq
Z(G,q,v) = \sum_{G^\prime \subseteq G} q^{k(G^\prime)}v^{e(G^\prime)} \ .
\label{cluster}
\label{zpol}
\eeq
This formula enables one to generalize $q$ from ${\mathbb Z}_+$ to ${\mathbb
R}$ or, indeed, ${\mathbb C}$.  From it one also directly infers the
equivalence 
\beq
Z(G,q,v)=(x-1)^{k(G)}(y-1)^{n}T(G,x,y)
\label{zt}
\eeq
where
\beq
x=1+\frac{q}{v}
\label{xqv}
\eeq
and
\beq
y=v+1
\label{yqv}
\eeq
so that
\beq
q=(x-1)(y-1) \ .
\label{qxy}
\eeq
Combining (\ref{ft}) and (\ref{zt}), one has the relation 
\beq
F(G,q) = (-1)^{|E|}q^{-n}Z(G,q,-q) \ .
\label{fz}
\eeq
Thus, the flow polynomial is a special case of the $q$-state Potts model given,
up to the above prefactor, by the evaluation
\beq
v=-q \ , \quad i.e., \quad K = \ln(1-q) \ .
\label{vqm}
\eeq
For the usual case of flows with $q \ge 2$, the condition $v=-q$
corresponds to a complex-temperature regime for the Potts model.  In
eq. (\ref{vqm}), one would thus have $K=\ln(q-1)+(2\ell+1)i\pi$, $\ell \in
{\mathbb Z}$.

\section{Some General Properties of Flow Polynomials} 

\subsection{Duality Relation with Chromatic Polynomial}

Before presenting our new results, we recall some basic information about flow
polynomials that will be relevant for our work.  We first discuss a useful
connection with chromatic polynomials.  The chromatic polynomial $P(G,q)$ of a
graph $G$ counts the number of proper $q$-coloring of $G$, where a proper
$q$-coloring is a coloring of the vertices of the graph using $q$ colors,
subject to the condition that the colors assigned to adjacent vertices are
different.  The minimum (positive integral) value of $q$ that allows one to
carry out a proper $q$-coloring of $G$ is called the chromatic number of $G$,
$\chi(G)$. The chromatic polynomial is given in terms of the Potts model
partition function and Tutte polynomial as
\beq
P(G,q) = Z(G,q,v=-1) = (-1)^{n+1}q^{k(G)}T(G,x=1-q,y=0) \ .
\label{pzt}
\eeq
Physically, the chromatic polynomial is the special case of the partition
function for the Potts antiferromagnet at zero temperature, $v=-1$.  Now let
$G$ be a planar graph and $G^*$ its planar dual graph.  Then
\beq
F(G,q)=q^{-1}P(G^*,q) \ .
\label{fpdual}
\eeq
This follows directly from the expressions for the flow and
chromatic polynomials in terms of the Tutte polynomial, together with the
symmetry relation of a Tutte polynomial
\beq
T(G,x,y) = T(G^*,y,x)
\label{tdual}
\eeq
for a planar graph $G$.  The equality (\ref{fpdual}) will be important for our
later discussion.  Several corollaries follow from this equality.  First, 
let $G$ be a bridgeless planar graph and $G^*$ its planar dual, which thus has
no loops.  Then 
\beq
\phi(G)=\chi(G^*) \ .
\label{phichi}
\eeq
(The restriction that $G$ has no bridges is made so that flows exist on
$G$, and this is dual to the restriction that $G^*$ has no loops, so that 
proper $q$-colorings of $G^*$ exist.)

\subsection{Some Results on Existence of $q$-Flows}
\label{qexistence}

We mention here a few mathematical results on existence of $q$-flows that are
relevant to our work.  Let us denote the degree or coordination number of a
vertex in a graph $G$ (i.e., the number of edges connected to this vertex) as
$\Delta$.  An elementary theorem states that a bridgeless graph admits a 2-flow
if and only if all of its vertex degrees are even.  This can be understood as
follows.  After choosing an orientation for the edges, assign to each oriented
edge the flow value 1.  Since $1 = -1$ mod 2, and since $2n = 0$ mod 2, the
fact that each vertex has even degree means that the flows into each vertex are
0 mod 2. A related statement is that if a graph has flow number $\phi(G)=2$,
then, since (for the nowhere-zero flows considered here) there is only one
choice of flow number for each edge, namely 1, which is the same as $-1$ mod 2,
it follows that $F(G,q=2)=1$. 

One of the most important theorems pertaining to flow polynomials for planar
graphs is the 4-flow theorem, which states that every bridgeless planar graph
has a 4-flow.  This is equivalent to the celebrated 4-color theorem
\cite{ah,ahk}, that every bridgeless planar graph has a coloring of faces with
four colors such that any two faces that are adjacent across a given edge have
different colors. An equivalent expression of the theorem for proper vertex
colorings is that every loopless planar graph has a proper 4-coloring.  One can
inquire about the existence of flows for arbitrary, not necessarily planar,
graphs.  Jaeger proved that every bridgeless graph (not necessarily planar) has
an 8-flow \cite{jaeger}, and subsequently Seymour proved the stronger result
that every bridgeless graph has a 6-flow \cite{seymour81}.  An outstanding
conjecture, due to Tutte \cite{tutte54}, is that every bridgeless graph has a
5-flow.  These general results provide a useful background for the specific
flow numbers that we shall obtain for various planar and nonplanar families of
graphs.

\subsection{Structural Properties for General Strip Graphs}

Since a flow polynomial is a special case of a Tutte polynomial or
equivalently, a Potts model partition function, one can obtain several
properties of flow polynomials from corresponding properties of the latter two
polynomials.  A general form for the Tutte polynomial for the strip graphs
considered here, or more generally, for recursively defined families of graphs
$G_m$ comprised of $m$ repeated subunits (e.g. the columns of squares of height
$L_y$ vertices that are repeated $L_x=m$ times to form strip of a regular
lattice of width $L_y$ and length $L_x$ (denoted $L_y \times L_x$) with some
specified boundary conditions), is \cite{a}
\beq 
T(G_m,x,y)=\frac{1}{x-1}\sum_{j=1}^{N_{T,G,\lambda}} c_{G,j}(\lambda_{T,G,j})^m
\label{tgsum}
\eeq
where the terms $\lambda_{T,G,j}$, the coefficients $c_{G,j}$, and the total
number $N_{T,G,\lambda}$ depend on $G$ through the type of lattice, its width,
$L_y$, and the boundary conditions, but not on the length.  Equivalently, 
\beq
Z(G_m,q,v)=\sum_{j=1}^{N_{Z,G,\lambda}} c_{G,j}(\lambda_{Z,G,j})^m
\label{zgsum}
\eeq
where 
\beq
N_{Z,G,\lambda} = N_{T,G,\lambda} \ .
\label{nztotnttot}
\eeq
It follows from (\ref{ft}) and (\ref{tgsum}) that the flow polynomial for
recursive families of graphs, comprised of $m$ repeated subgraph units, has the
general form
\beq
F(G_m,q) = \sum_{j=1}^{N_{F,G,\lambda}} c_{G,j} (\lambda_{F,G,j})^m
\label{fgsum}
\eeq
where again the terms $\lambda_{F,G,j}$, the coefficients $c_{G,j}$, and the 
total number of terms $N_{F,G,\lambda}$ depend on $G$ through the type of 
lattice, its width, $L_y$, and the boundary conditions, but not on the length.
This is related to (\ref{tgsum}) as follows.  The prefactor in (\ref{ft}),
$(-1)^{|E|-|V|+1}$ is of the form $(-1)^{cm+1}$, where $c$ is an even or odd
integer.  The single factor $(-1)$ cancels the $(-1)$ factor resulting from the
$x=0$ evaluation of the overall prefactor $1/(x-1)$ in (\ref{tgsum}).  For the
cyclic lattice strips considered here, both $|V|$ and $|E|$ are integer
multiples of $L_x=m$, so their difference is of the stated form, $cm$.  Thus,
\beq
\lambda_{F,G,j}(q) = (-1)^c \lambda_{T,G,j}(x=0,y=1-q)
\label{flam_lamtut}
\eeq
for the subset of the $\lambda_{T,G,j}$'s that are nonzero when evaluated at
$x=0$.

For a given type of strip graph, the sum of the coefficients in (\ref{fgsum})
is denoted 
\beq
C_{F,G} = \sum_{j=1}^{N_{F,G,\lambda}} c_{G,j} \ .
\label{csum}
\eeq

\subsection{The Function $fl$ and Associated Locus ${\cal B}$ }

Given the structural property (\ref{fgsum}), it is natural, in the limit $|V|
\to \infty$, to define a function specifying the number of flows per face,
\beq
fl(\{G\},q) = \lim_{|V| \to \infty} F(G,q)^{1/f(G)}
\label{fl}
\eeq
For almost all of the strip graphs considered here, $|V| \to \infty \
\Longleftrightarrow \ f(G) \to \infty$.  An exception is the circuit graph, for
which $F(G)=2$, independent of $|V|$.

For reference, the reduced free energy is 
\beq
f(\{G\},q,v) = \lim_{|V| \to \infty} \ln [Z(G,q,v)^{1/|V|}]
\label{f}
\eeq
a limiting function obtained from the Tutte polynomial may be written as 
\beq
\tau(\{G\},x,y) = \lim_{|V| \to \infty} T(G,x,y)^{1/|V|}
\label{tau}
\eeq
and the ground state degeneracy per vertex of the $q$-state Potts
antiferromagnet is 
\beq
W(\{G\},q) = \lim_{|V| \to \infty} P(G,q)^{1/|V|} \ .
\label{w}
\eeq
Since $F(G,q)$ is a polynomial in $q$ of degree $c(G)$, it follows that with
this definition,
\beq
\lim_{q \to \infty} \frac{fl(\{G\},q)}{q} = 1 \ .
\label{flqlim}
\eeq
This is the analogue to the property that
\beq
\lim_{q \to \infty} \frac{W(\{G\},q)}{q} = 1 \ .
\label{wqlim}
\eeq

Following our earlier nomenclature \cite{w}, we denote a term $\lambda$ in
(\ref{fgsum}) as leading (= dominant) for a given value of $q$ if it has a
magnitude greater than or equal to the magnitude of other $\lambda$'s evaluated
at this value of $q$.  In the limit $|V| \to \infty$ the leading $\lambda$ in
$F(G_m,q)$ determines the function $fl(\{G\},q)$.  The continuous locus ${\cal
B}$ where $fl(\{G\},q)$ is nonanalytic thus occurs where there is a switching
of dominant $\lambda$'s in $F$ and is the solution of the equation of
degeneracy in magnitude of these dominant $\lambda$'s.  Since a zero in
$F(G,q)$ requires a cancellation between dominant $\lambda$'s, it is clear that
as $|V| \to \infty$, the continuous accumulation set ${\cal B}$ of the zeros of
$F(G,q)$ forms the boundary curves across which this switching occurs and hence
across which $fl(\{G\})$ is nonanalytic.  Depending on the family of strip
graphs, the locus ${\cal B}$ may or may not separate the complex $q$ plane into
different regions and may or may not cross the real $q$ axis.  If it does, we
denote the maximal point where it crosses this axis as $q_{cf}(\{G\})$.  We
also denote the region including the positive real $q$ axis extending down from
$q=\infty$ (and terminating at $q_{cf}(\{G\})$ if the latter point exists) as
region $R_1$.  This region $R_1$ is understood to include the maximal area to
which one can analytically continue $fl(\{G\})$ from the large-$q$ positive
real axis.  For families of strip graphs where ${\cal B}$ separates the $q$
plane into different regions, the function $fl(\{G\})$ has different analytic
forms in the different regions.  For a strip graph $G$, in region $R_1$, 
\beq
fl(\{G\},q) = \lambda_{R1,dom.}
\label{flr1}
\eeq
where $\lambda_{R1,dom.}$ denotes the dominant $\lambda$ in region $R_1$.

Just as was true for these functions $f$, $\tau$, and $W$, there are two
subtleties in the definition (\ref{fl}): (i) which of the $c(G)$ roots to take
in the equation, and (ii) the fact that at certain values of $q$, one has the
noncommutativity of limits
\beq
\lim_{|V| \to \infty} \lim_{q \to q_s} F(G,q)^{1/f(G)} \ne
\lim_{q \to q_s} \lim_{|V| \to \infty} F(G,q)^{1/f(G)} \ .
\label{flnoncom}
\eeq
We have discussed these before in the context of the definitions of $W$ and $f$
\cite{w,a,bcc}.  Concerning item (i), for sufficiently large real $q$, in the
region $R_1$, $F(G,q)$ is real and positive, so one chooses the canonical root
for $fl(\{G\})$, which is real and positive.  However, if ${\cal B}$ separates
the $q$ plane into different regions, then in the latter regions, there is no
canonical choice of phase that one can make for the root (\ref{fl}) and hence
only the magnitude $|fl(\{G\},q)|$ can be determined unambiguously.  For item
(ii) the noncommutativity typically occurs at special values 
$q_s=1,...,\phi(G)-1$. To avoid isolated discontinuities that would otherwise
result where $fl$ vanishes, we shall, following our earlier practice for $W$
and $f$ \cite{w,a} and adopt, for the cases where such noncommutativity occurs,
the second order of limits in (\ref{flnoncom})
\beq
fl(\{G\},q_s) = \lim_{q \to q_s} \lim_{|V| \to \infty} F(G,q)^{1/f(G)} \
.
\label{flqs}
\eeq

 From the duality relation (\ref{fpdual}) it follows that if $G$ is a planar
graph and $G^*$ is its planar dual, then, taking into account that the number
of faces of $G$ is equal to the number of vertices of $G^*$, one has 
\beq
fl(\{G\},q) = W(\{G^*\},q) \ .
\label{flw}
\eeq
Note that if we had defined $fl(\{G\},q)$ as $\lim_{|V| \to \infty}
F(G,q)^{1/|V|}$, we would have obtained a different relation $fl(\{G\},q) =
W(\{G^*\},q)^p$, where $p$ is a geometric factor depending on the particular
graphs $G$.  Since the infinite square lattice is self-dual, it follows that
\beq
fl(sq,q) = W(sq,q) \ .
\label{flsqwsq}
\eeq
Since the infinite honeycomb and triangular lattices are the duals of each
other, we also have 
\beq
fl(tri,q)=W(hc,q) \ , \quad fl(hc,q)=W(tri,q) \ .
\label{fltriwsq}
\eeq
The dual of the kagom\'e lattice is the diced lattice \cite{gs,wn}, and hence 
\beq
fl(kag,q)=W(diced,q) \ , \quad fl(diced,q)=W(kag,q) \ .
\label{flkagdiced}
\eeq

We next give some results for the locus ${\cal B}$.  Because of the fact that
these $\lambda$'s are degenerate in magnitude on ${\cal B}$, it follows
that $fl(\{G\},q)$ is nonanalytic but continuous across ${\cal B}$.  A
basic property is that ${\cal B}(\{G\})$ is invariant under complex
conjugation,
\beq
{\cal B}(\{G\}) = {\cal B}(\{G\})^* \ .
\label{bbstar}
\eeq
This is a consequence of the property that the coefficients of each term in the
flow polynomial are real and hence the zeros of the flow polynomial are
invariant under complex conjugation.  Hence, the same property holds for their
asymptotic accumulation set as $|V| \to \infty$.

We next point out another consequence of (\ref{fpdual}) and (\ref{flw}) for
planar graphs.  For this purpose, we must append subscripts to distinguish
${\cal B}_{fl}$ and ${\cal B}_{W}$, the respective continuous accumulation sets
of zeros of the flow and chromatic polynomials in the limit $|V| \to \infty$.
Then for the $|V| \to \infty$ limits of planar graphs $\{G\}$,
\beq
{\cal B}_{fl}(\{G\}) = {\cal B}_{W}(\{G^*\}) \ .
\label{bflwdual}
\eeq
Several corollaries follow.  Since the infinite square lattice is self-dual, 
\beq
{\cal B}_{fl}(sq) = {\cal B}_W(sq)
\label{bflwsq}
\eeq
Since the infinite triangular and honeycomb lattices are dual to each other,
we have
\beq
{\cal B}_{fl}(tri) = {\cal B}_W(hc) \ , \quad 
{\cal B}_{fl}(hc) = {\cal B}_W(tri)
\label{bflwtrihc}
\eeq
Since the kagom\'e and diced lattices (\cite{gs,wn}) are dual to each other, 
\beq
{\cal B}_{fl}(kag) = {\cal B}_W(diced) \ , \quad
{\cal B}_{fl}(diced) = {\cal B}_W(kag)
\label{bflwkagdiced}
\eeq

Now consider the lattice strip graphs $G[L_y \times L_x,cyc]$ and $G[L_y \times
L_x,Mb.]$. We have shown earlier \cite{pm,bcc,tor4} that the locus ${\cal B}_W$
is the same for the respective infinite-length limits of both of these strip
graphs.  We also find that the same property holds for the ${\cal B}_{fl}$
computed here, but note that this does not follow from our previous results for
${\cal B}_W$ since the M\"obius strips are nonplanar, and hence one cannot
apply the duality relation (\ref{fpdual}) and the consequent equalities
(\ref{flw}) and (\ref{bflwdual}).  We also find that for the families of strip
graphs that we have calculated, ${\cal B}_{fl}$ is the same for boundary
conditions corresponding to embedding on surfaces with torus and Klein bottle
topologies.

\subsection{Noncompactness of ${\cal B}$ for Classes of Strip Graphs}

An important feature of the loci ${\cal B} \equiv {\cal B}_{fl}$ for the
infinite-length limits of the families of lattice strips with periodic (or
twisted periodic) longitudinal boundary conditions is that many of these are
noncompact in the $q$ plane, containing curves that extend infinitely far away
from the origin.  {\it A fortiori}, this means that for this families of
graphs, as $L_x \to \infty$, there is no upper bound on the magnitudes $|q|$ of
the zeros of the flow polynomial.  The noncompactness of ${\cal B}_{fl}$ for
many strips of regular lattices is quite different from the loci ${\cal B}_W$
for the corresponding strips, which, as we showed in our previous works on
chromatic polynomials, were compact in the $q$ plane \cite{w}-\cite{s5}. Sokal
has proved that for an arbitrary graph $G$, any zero of the chromatic
polynomial satisfies the following upper bound \cite{sokalzero}
\beq
P(G,q)=0 \ \Longrightarrow \ |q| \le c \Delta, \quad c \simeq 7.964
\label{sokal}
\eeq
where here $\Delta$ denotes the maximal vertex degree in $G$. 
(Indeed, our explicit calculations for a variety of families of strip graphs
yielded loci ${\cal B}_W$ on which the values of ${\rm max}(|q|)$ were
considerably less than the above-mentioned upper bounds for the respective
strips.)  As was discussed in \cite{wa,wa2,wa3}, the locus ${\cal B}_W$ is
noncompact if and only if the degeneracy condition in magnitude of two or more
different dominant $\lambda$'s can be satisfied for arbitrarily large $|q|$.
Clearly, the same condition holds for ${\cal B}_{fl}$.  One of us (RS), with
Tsai, studied the conditions under which this could occur and constructed
several general classes of families of graphs that were designed to satisfy
this condition and thereby yield noncompact loci ${\cal B}_W$
\cite{w,wa,wa2,wa3} (see also \cite{read91,sokalzero}).  As noted in
\cite{wa,wa2}, the condition that $\lim_{|V| \to \infty} \Delta = \infty$ is a
necessary but not sufficient condition for ${\cal B}_W$ to be noncompact; an
example of a graph with a maximal vertex degree that goes to infinity as $|V|
\to \infty$ but with a compact ${\cal B}_W$ is the wheel graph $K_1 + C_m$.
Here, $C_m$ denotes the circuit graph, $K_p$ denotes the complete graph on $p$
vertices, defined as the graph each of whose vertices is adjacent to every
other vertex, and the ``join'' $G+H$ is defined as the graph formed by
connecting each of the vertices of $G$ to each of the vertices of $H$. (The
${\cal B}$ in this case is simply the unit circle $|q-2|=1$.)

Using the duality relation (\ref{fpdual}) for planar graphs, one can understand
why ${\cal B}_{fl}$ is noncompact for the simplest nontrivial cyclic strip of
the square lattice, namely for width $L_y=2$.  To do this, we observe that the
planar dual of this $2 \times L_x$ cyclic strip is the circuit graph with $L_x$
vertices, augmented by having an extra vertex connected by edges to each of the
vertices on the upper side of the strip and, separately, an extra vertex
connected by edges to each of the vertices on the lower side of the strip.
This is a special case of one of the families of graphs that were shown in
\cite{wa,wa2} to yield, in the $L_x \to \infty$ limit, noncompact loci ${\cal
B}$, namely the family $(K_p)_b+ G_r$, where, $b$ means that $b$ of the edges
in the complete graph $K_p$ are cut, and $G_r$ denotes an arbitrary $r$-vertex
graph. We refer the reader to section II of \cite{wa} and section 2 of
\cite{wa2} for further details.  Specifically, we find that the planar dual
graph to the $L_y=2$ cyclic stripf of the square lattice is the special case of
$(K_p)_b+ G_r$ with $p=2$, $b=1$, signifying the cutting of the single edge in
the $K_2$, and $G_r=C_{L_x}$, the circuit graph.  Using the duality relation
(\ref{bflwdual}), we therefore prove that ${\cal B}_{fl}$ for the $L_x \to
\infty$ limit of the cyclic $L_y \times L_x$ strip is noncompact, extending
infinitely far from the origin in the $q$ plane.  Similar arguments give
insight into the noncompactness of ${\cal B}$ for the wider strips considered
here. Because of the noncompactness of ${\cal B}_{fl}$ in the $q$ plane and the
property that we find that this locus does not pass through $q=0$, it will
often be convenient to display it in the plane of the inverse variable
\beq
u = \frac{1}{q}
\label{u}
\eeq
where it is compact.

We note that not all strip graphs with periodic longitudinal boundary
conditions lead to noncompact ${\cal B}_{fl}$.  An example of a family with
compact ${\cal B}_{fl}$ is provided by the cyclic self-dual strips of the
square lattice $G_D(L_y \times L_x)$ that we studied in \cite{dg,sdg}.  Since
these are self-dual planar graphs, one can immediately use (\ref{bflwdual}) to
infer that the loci ${\cal B}_{fl}$ are identical to the loci ${\cal B}_W$
given in \cite{dg,sdg}, which are compact.

In contrast to the case with many families of strip graphs with periodic (or
twisted periodic) longitudinal boundary conditions, we find that ${\cal B}$ is
compact for the $L_x \to \infty$ limit of strips with free longitudinal
boundary conditions, as will be illustrated with specific exact solutions
below.  This feature is similar to our previous findings for ${\cal B}_W$.
Henceforth, where no confusion will result, we shall drop the subscript $fl$ on
${\cal B}_{fl}$.

\section{General Structural Results for Cyclic Strips of the Square and
Honeycomb Lattices}

In \cite{cf} it was shown that for cyclic and M\"obius strips of the square
lattice of fixed width $L_y$ and arbitrary length $L_x$ (and also for cyclic
strips of the triangular lattice) the coefficients $c_j$ in the Tutte
polynomial are polynomials in $q$ with the property that for each degree $d$
there is a unique polynomial, denoted $c^{(d)}$.  Further, this was shown to be
\beq
c^{(d)} = U_{2d}(q^{1/2}/2) = \sum_{j=0}^d (-1)^j {2d-j \choose j} 
q^{d-j}
\label{cd}
\eeq
where $U_n(x)$ is the Chebyshev polynomial of the second kind.  A number of
properties of these coefficients were derived in \cite{cf}.  
We list below the specific $c^{(d)}$'s that will be needed here: 
\beq
c^{(0)}=1 \ , \quad c^{(1)}=q-1 \ , \quad c^{(2)}=q^2-3q+1 \ ,
\label{cd012}
\eeq
\beq
c^{(3)}=q^3-5q^2+6q-1 \ .
\label{cd3}
\eeq

Thus, the terms $\lambda_{T,L_y,j}$ that occur in (\ref{tgsum}) can be
classified into sets, with the $\lambda_{T,L_y,j}(q,v)$ in the $d$'th set being
defined as those terms with coefficient $c^{(d)}$.  In Ref. \cite{cf} the
numbers of such terms, denoted $n_T(L_y,d)$, were calculated.  Labelling the
eigenvalues with coefficient $c^{(d)}$ as $\lambda_{T,L_y,d,j}$ with $1 \le j
\le n_T(L_y,d)$, the Tutte polynomial for a cyclic strip graph of length
$L_x=m$ can be written in the form \cite{cf} 
\beq
T(G_s[L_y \times m; cyc.],x,y) = \frac{1}{x-1}\sum_{d=0}^{L_y} 
c^{(d)} \sum_{j=1}^{n_T(L_y,d)} (\lambda_{T,G_s,L_y,d,j})^m \ .
\label{tgsumcyc}
\eeq
For the M\"obius strip of the square lattice the coefficients may be either
$+c^{(d)}$ or $-c^{(d)}$ and the terms are accordingly labelled as
$\lambda_{T,L_y,d,\pm,j}$, where $1 \le j \le n_T(L_y,d,\pm)$. We have
\cite{cf}, using the notation $TPBC_y$ for twisted periodic b.c. in the
longitudinal direction,
\beq
T(sq[L_y \times m;Mb.],x,y) = \frac{1}{x-1}\sum_{d=0}^{d_{max}}
c^{(d)}\biggl [\sum_{j=1}^{n_T(L_y,d,+)} (\lambda_{T,L_y,d,+,j})^m
-\sum_{j=1}^{n_T(L_y,d,-)} (\lambda_{T,L_y,d,-,j})^m \biggr ]
\label{tgsummb}
\eeq
where
\beq
d_{max}= \cases{\frac{L_y}{2} & for \ even \ $L_y$ \cr\cr
\frac{L_y+1}{2} & for \ odd \ $L_y$} \ .
\label{dmaxmb}
\eeq
The number $n_T(G_s,L_y,d)$ of $\lambda$'s with a given coefficient $c^{(d)}$
is \cite{cf}
\beq 
n_T(G_s,L_y,d)=\frac{(2d+1)}{(L_y+d+1)}{2L_y \choose L_y-d} \quad {\rm
for} \ \ G_s=sq,tri,hc; \quad 0 \le d \le L_y
\label{ntlyd}
\eeq
and zero otherwise.  The total number $N_{T,L_y,\lambda}$ of
different terms $\lambda_{T,L_y,j}$ in eq. (\ref{tgsum}) for cyclic (or
M\"obius) strips $G_s$ of the square, triangular, and honeycomb lattices is
\cite{cf}
\beq
N_{T,G_s,L_y,\lambda}=\sum_{d=0}^{L_y} n_T(G_s,L_y,d)
\label{nztot}
\eeq
which was calculated to be \cite{cf,hca}
\beq
N_{T,G_s,L_y,\lambda}={2L_y \choose L_y} \ .
\label{nztotcyc}
\eeq
For arbitrary $L_y$, eq. (\ref{ntlyd}) shows that there is a unique 
$\lambda_{T,L_y,d}$ corresponding to the coefficient $c^{(d)}$ of highest 
degree, $d=L_y$, and this term is
\beq
\lambda_{T,G_s,L_y,d=L_y} = 1 \ .
\label{lamlyly}
\eeq
Hence, 
\beq
\lambda_{F,sq,L_y,d=L_y}=\lambda_{F,hc,L_y,d=L_y}=(-1)^{L_y+1}
\label{fsqcly}
\eeq
and
\beq
\lambda_{F,tri,L_y,d=L_y}=1
\label{ftricly}
\eeq
(independent of $L_y$).  Since this eigenvalue is unique, it is not necessary
to append another index, as with the other $\lambda$'s, and we avoid this for
simplicity.

 From (\ref{tgsumcyc}), it follows that the flow polynomial for a cyclic strip
of the $G_s$-type lattice also has the same type of form, i.e., 
\beq
F(G_s[L_y \times m; cyc.],x,y) = \sum_{d=0}^{L_y}
c^{(d)} \sum_{j=1}^{n_F(G_s,L_y,d)} (\lambda_{F,G_s,L_y,d,j})^m \ .
\label{fgsumcyc}
\eeq
Let us denote the number of $\lambda_{F,G_s,L_y,d,j}$ with coefficient 
$c^{(d)}$ as $n_F(G_s,L_y,d)$.  Then for this strip 
\beq
C_{F,G_s,L_y}=\sum_{d=0}^{L_y} n_F(G_s,L_y,d)c^{(d)} \ .
\label{fcsumcyc}
\eeq
We now concentrate on cyclic strips of the square and honeycomb lattice.  Aside
from degenerate cases, the dual of the cyclic strip of the square lattice of
width $L_y \ge 2$ and length $L_x$ is a strip of the same lattice with width
$L_y-1$ and length $L_x$ with all of the vertices along the upper and lower
sides connected to two respective external vertices.  Now consider proper
colorings of this dual graph.  The total number of proper $q$-colorings of a
transverse slice is $q(q-1)^{L_y}$.  It is elementary to show that this also
holds for $L_y=1$. From (\ref{fpdual}), it follows that 
\beq
C_{F,sq,L_y} = (q-1)^{L_y} \ .
\label{ncrelcyc}
\eeq
A similar argument shows that (\ref{ncrelcyc}) also holds for cyclic strips 
of the honeycomb lattice.

\bigskip

We now derive the following structural theorem.

\begin{th} \quad Consider the flow polynomial for a cyclic strip graph $G_s$ 
of the square or honeycomb lattice of fixed width $L_y$ and arbitrarily great 
length $L_x$.  For brevity, set $n_F(G_s,L_y,d) \equiv n_F(L_y,d)$. 
The $n_F(L_y,d)$, $d=0,1,..L_y$ are determined as follows. One has
\beq
n_F(L_y,d)=0 \quad {\rm for} \quad d > L_y \ ,
\label{nfup}
\eeq
\beq
n_F(L_y,L_y)=1
\label{nfcly}
\eeq
\beq
n_F(1,0)=0
\label{nfly10}
\eeq
with all other numbers $n_F(L_y,d)$ being determined by the two recursion
relations
\beq
n_F(L_y+1,0)=n_F(L_y,1)
\label{nfrecursion1}
\eeq
and
\beq
n_F(L_y+1,d) = n_F(L_y,d-1)+n_F(L_y,d)+n_F(L_y,d+1)
\quad {\rm for} \quad L_y \ge 1 \quad {\rm and} \quad 1 \le d \le L_y+1 \ .
\label{nfrecursion2}
\eeq

\label{ncalftheorem}
\end{th}

\bigskip

Proof. \  We substitute for $c^{(d)}$ from eq. (\ref{cd}) in eq.
(\ref{ncrelcyc}).  We obtain another equation by differentiating this with
respect to $q$ once; another by differentiating twice, and so forth up to
$L_y$-fold differentiations.  This yields $L_y+1$ linear equations in the
$L_y+1$ unknowns, $n_F(L_y,d)$, $d=0,1,...,L_y$.  We solve this set of
equations to get the $n_F(L_y,d)$.  $\Box$

\begin{cor}
\beq
n_F(L_y,L_y-1)=L_y-1 \ .
\label{nfclyminus1}
\eeq

\label{nflylym1corollary}
\end{cor}

\begin{table}
\caption{\footnotesize{Table of numbers $n_F(L_y,d)$ and their sums,
$N_{F,L_y,\lambda}$ for cyclic strips of the square and honeycomb lattices. 
Blank entries are zero.}}
\begin{center}
\begin{tabular}{|c|c|c|c|c|c|c|c|c|c|c|c|c|}
\hline\hline
$L_y \ \downarrow$ \ \ $d \ \rightarrow$
   & 0 & 1   & 2   & 3   & 4   & 5  & 6  & 7 & 8 & 9 & 10 &
$N_{F,L_y,\lambda}$
\\ \hline\hline
1  & 0   & 1   &     &     &     &    &    &   &   &    &   & 1    \\ \hline
2  & 1   & 1   & 1   &     &     &    &    &   &   &    &   & 3    \\ \hline
3  & 1   & 3   & 2   & 1   &     &    &    &   &   &    &   & 7    \\ \hline
4  & 3   & 6   & 6   & 3   & 1   &    &    &   &   &    &   & 19   \\ \hline
5  & 6   & 15  & 15  & 10  & 4   & 1  &    &   &   &    &   & 51   \\ \hline
6  & 15  & 36  & 40  & 29  & 15  & 5  & 1  &   &   &    &   & 141  \\ \hline
7  & 36  & 91  & 105 & 84  & 49  & 21 & 6  & 1 &   &    &   & 393  \\ \hline
8  & 91  & 232 & 280 & 238 & 154 & 76 & 28 & 7 & 1 &    &   & 1107 \\ \hline
9  & 232 & 603 & 750 & 672 & 468 & 258& 111& 36& 8 & 1  &   & 3139 \\ \hline
10 & 603 & 1585&2025 &1890 &1398 & 837& 405&155& 45& 9 & 1  & 8953 \\
\hline\hline
\end{tabular}
\end{center}
\label{nfctablecyc}
\end{table}

We note that the recursion relations (\ref{nfrecursion1}), (\ref{nfrecursion2})
for the numbers $n_F(L_y,d)$ for cyclic strips of the square and honeycomb
lattices are the same as the recursion relations that we derived earlier in
eqs. (3.14) and (3.15) of \cite{cf} for the corresponding numbers $n_P(L_y,d)$
for cyclic strips of the square and triangular lattice.  Thus, the differences
in the values of $n_F(L_y,d)$ for cyclic strips of the square and honeycomb
lattice and the $n_P(L_y,d)$ for cyclic strips of the square and triangular
lattice are due to the different initial values of these quantities, i.e.
(\ref{nfly10}) in the former case and $n_P(1,0)=n_P(1,1)=1$ in the latter case.

\begin{cor}
\beq
n_F(L_y,0) = \frac{1}{L_y+1}\sum_{j=1}^{[(L_y+1)/2]} {L_y+1 \choose j}
{L_y-j-1 \choose j-1} \quad {\rm for} \quad L_y \ge 2 \ .
\label{nfly0}
\eeq

\label{nfly0corollary}
\end{cor}

\bigskip

Proof.  This follows immediately from the solution to our general recursion
relations (\ref{nfrecursion1}), (\ref{nfrecursion2}). $\Box$. 

\bigskip

Summing the $n_F(L_y,d)$ over $d$ for a given strip with $L_y$, we obtain 
\beq
N_{F,L_y,\lambda}=\sum_{j=0}^{[L_y/2]} {L_y \choose j}{L_y-j \choose j}
\label{nftot}
\eeq
where $[x]$ is integer part of $x$.  This total number also applies to M\"obius
strips of the square and honeycomb lattice. 

A generating function for the $N_{F,L_y,\lambda}$ is \cite{sl,bernhart} 
\beq
\frac{1}{\sqrt{1-2x-3x^2}} - 1 = \sum_{L_y=1}^\infty
N_{F,L_y,\lambda}x^{L_y} \ .
\label{nfgenfun}
\eeq
 From this, it follows that the number $N_{F,L_y,\lambda}$ grows
exponentially fast with the width $L_y$ of the cyclic strip of
the square or honeycomb lattice, with the leading asymptotic behavior
\beq
N_{F,L_y,\lambda} \sim L_y^{-1/2} \ 3^{L_y} \quad {\rm as} \ \ L_y \to \infty
 \ .
\label{nfasymp}
\eeq
Analogous structural results can also be given for cyclic strips of the
triangular lattice. We proceed to present exact calculations of flow
polynomials for a number of different lattice strips.

\section{Cyclic and M\"obius Strips of the Square Lattice}

\subsection{General} 

We denote the cyclic family as $sq(L_y,L_x,FBC_x,PBC_y) \equiv
sq(L_y,L_x,cyc.)$ and the M\"obius family as $sq(L_y,L_x,FBC_x,TPBC_y) \equiv
sq(L_y,L_x,Mb.)$.  The cyclic and M\"obius strips of the square lattice of
width $L_y$ and length $L_x$ have $|V|=L_xL_y$ vertices and $|E|=L_x(2L_y-1)$,
so that the prefactor in (\ref{ft}) is $(-1)^{L_x(L_y-1)+1}$ and hence the
$\lambda_{F,sq,L_y,d,j}$ are given by the subset of the
$\lambda_{T,sq,L_y,d,j}$'s that are nonzero when evaluated for $x=0$ and
$y=1-q$, multiplied by the prefactor $(-1)^{L_y-1}$.  Among these families of
graphs, the lowest case, $L_y=1$, is just the circuit graph $C_m$, and the
result is elementary; the only $\lambda_{F,sq,j}$ is unity, and $F(C_m,q)=q-1$
(independent of $m$).  This polynomial has only a single zero, at $q=1$.  With
the definition (\ref{fl}), we obtain $fl(q)=\sqrt{q-1}$.  Henceforth, for
brevity of notation, where no confusion will result, we shall omit the $F$ in
$\lambda_{F,sq,L_y,d,j}$ and write this simply as $\lambda_{sq,L_y,d,j}$, and
similarly for other lattice types.

\bigskip

\subsection{$L_{\lowercase{y}}=2$ Cyclic and M\"obius Strips of the Square 
Lattice}

Using the duality relation (\ref{fpdual}), one has
\beq
F(sq[2 \times L_x=m,cyc.],q) = q^{-1}P((K_2)_1 + C_m,q)
\label{sqpxy2wa}
\eeq
where $K_p$ is the complete graph, $G+H$ is the join of the graphs $G$ and $H$,
$(K_p)_b$ denotes the graph obtained by removing $b$ edges from $K_p$, and
$C_m$ is the circuit graph with $m$ vertices.  In \cite{w,wa,wa2}, one of us
(RS), with Tsai, calculated the chromatic polynomial for the family $(K_2)_1 +
C_m$.  This immediately gives the flow polynomial, via eq. (\ref{sqpxy2wa}),
with
\beq
\lambda_{sq,2,0,1}=q-2
\label{flam201}
\eeq
\beq
\lambda_{sq,2,1,1}=q-3
\label{flam211}
\eeq
\beq
\lambda_{sq,2,2}=-1 \ .
\label{flam22}
\eeq
Hence, we find 
\beq
F(sq[2 \times m,cyc.],q)=(q-2)^m+c^{(1)}(q-3)^m+c^{(2)}(-1)^m \ .
\label{flowsqpxy2}
\eeq
Note that $n_F(2,0)=n_F(2,1)=n_F(2,2)=1$ and $N_{F,2,\lambda}=3$, in accord
with our general structural formulas given above. 

By using results that were obtained in \cite{a} for the Tutte polynomial,
we obtain 
\beq
F(sq[2 \times m, Mb.],q)=(q-2)^m+c^{(1)}(q-3)^m-(-1)^m \ .
\label{flowsqpxy2mb}
\eeq
Thus, 
\beq
n_F(sq,2,0,+)=1, \quad n_F(sq,2,0,-)=1, \quad n_F(sq,2,1,+)=1, 
\quad n_F(sq,2,1,-)=0
\label{nfsqpxy2dmb}
\eeq
with $n_F(sq,2,d,\pm)=0$ for $d \ge 2$. 

We observe that $F(sq[L_y=2,L_x=m,cyc.],q)$ and
$F(sq[L_y=2,L_x=m,Mb.],q)$ always have the factors $(q-1)(q-2)$.  For $m
\ge
3$ odd, $F(sq[L_y=2,L_x=m,cyc.],q)$ also has the factor $(q-3)$, while for
$m$ even, $F(sq[L_y=2,L_x=m,Mb.],q)$ also has the factor $(q-3)$.  
These results show that 
\beq
\phi(sq[2 \times m,cyc.])=\cases{3 & if $m$ is even \cr\cr
                                 4 & if $m \ge 3$ is odd}
\label{phisqpxy2}
\eeq
(for $m=1$, this flow polynomial vanishes since the graph contains a
bridge).  Further, 
\beq
\phi(sq[2 \times m,Mb.])=\cases{4 & if $m$ is even \cr\cr
                                3 & if $m$ is odd} 
\label{phisqpxy2mb}
\eeq

\bigskip

We recall that a graph $G$ is $k$-critical if $P(G,q=\chi(G))=k!$, where the
chromatic number of $G$, $\chi(G)$, was defined above.  Somewhat analogously to
the concept of $k$-critical graphs for proper vertex coloring and chromatic
polynomials, one may ask whether for a graph $G$ with $\phi(G)=k$ it is true
that $F(G,q=\phi(G))$ has a fixed value depending on $q$, i.e. for $q=\phi(G)$,
the number of flows on $G$ is a fixed number rather than growing
(exponentially) with $|V|$.  Using our results we can answer the question for
these cyclic and M\"obius strip graphs.  First, we recall the elementary
observation that if $\phi(G)=2$, then $F(G,2)=1$.  From eqs. (\ref{phisqpxy2})
and (\ref{phisqpxy2mb}) we know that the flow number is 3 for the $L_y=2$
cyclic and M\"obius strips of the square lattice with even and odd $L_x=m$,
respectively, and we find
\beq
F(sq[2 \times m,cyc.],q=3)=2 \quad {\rm for \ even} \ \  m \ge 2
\label{flowsqpxy2q3}
\eeq
and
\beq
F(sq[2 \times m,Mb.],q=3)=2  \quad {\rm for \ odd} \ \ m \ge 1 \ .
\label{flowsqtpx2q3}
\eeq
In contrast, it is readily verified from our exact results that for 
the $L_y=2$ cyclic and M\"obius strips of the square lattice with odd and even
$L_x=m$, respectively, for which $\phi=4$, the evaluation of the respective
flow polynomials at $q=4$ yields values that grow (exponentially) with the 
length of the strip.  

One way to prove eqs. (\ref{flowsqpxy2q3}) and (\ref{flowsqtpx2q3}) is to use
our explicit calculations of the flow polynomials for these families of graphs.
Another way is to write down the actual flows.  For this purpose, consider a
given square on the strip.  For $q=3$ there are two circular flows on this
square, namely those with a flow number 1 assigned to each edge, going in (i) a
clockwise or (ii) a counterclockwise manner.  Clearly, a flow with flow number
1 assigned to each edge, going in a clockwise manner is equivalent to a flow
with flow number $2=-1$ mod 3 assigned to each edge going in a counterclockwise
manner.  Consider the flow of type (i) on this square.  In order to satisfy the
flow conservation condition at the right-hand upper and lower vertices, it is
necessary that the flow on the neighboring square to the right be of type (ii).
Similarly, if the flow on the given square is of type (ii), then it is
necessary that the flow on the neighboring square to the right be of type (i).
Continuing in this manner along the cyclic strip, one finds a consistent set of
choices if $m$ is even, but not if $m$ is odd.  This proves the first part of
the corollary. A similar proof with obvious changes works for the second part.

\bigskip

We remark that the constructive proofs given above establish that for $q=3$ the
total flows can be decomposed into superpositions of circulations around each
square.  Since there is a 1-1 correspondence with the flows around a face of a
graph and the face-colorings of the graph \cite{tutte54}, one sees that these
corollaries are equivalent to an analogous condition for the face-coloring of
the strip graph, or equivalently, the vertex coloring of the dual graph. 

\bigskip

In accordance with our general discussion given above, the locus ${\cal B}$ for
the $L_x \to \infty$ limit of the cyclic and M\"obius $L_y=2$ strips of the
square lattice is noncompact in the $q$ plane.  From eq. (\ref{sqpxy2wa}), it
follows that this locus is identical to the one that was determined earlier for
the $m \to \infty$ limit of the family $P((K_2)_1 + C_m,q)$ in \cite{w,wa,wa2}
and shown in the $q$ plane shown in Fig. 2 of \cite{w}, or equivalently, to the
locus ${\cal B}$ in the $u=1/q$ plane shown in Fig. 1 of \cite{wa2}.  It
divides the $q$ plane into three regions, $R_i$, $i=1,2,3$.  Regions $R_1$,
$R_2$, and $R_3$ contain the respective real intervals $3 \le q \le \infty$; $2
\le q \le 3$, and $q \le 2$.  Region $R_2$ is bounded on the left by the arc of
the circle
\beq
q = 3 + e^{i\theta} \ , \quad {\rm for} \quad 
\frac{2\pi i}{3} \le \theta \le \frac{4\pi i}{3}
\label{arcleft}
\eeq
and on the right by the arc of the circle 
\beq
q = 2 + e^{i\theta} \ , \quad {\rm for} \quad
 -\frac{\pi i}{3} \le \theta \le \frac{\pi i}{3} \ .
\label{arcright}
\eeq
Thus,
\beq
q_{cf} = 3 \quad {\rm for} \quad \{G\} = sq, \ 2 \times \infty, \ cyc./Mb.
\label{qcsqpxy2}
\eeq
The two arcs (\ref{arcleft}) and (\ref{arcright}) meet at the complex-conjugate
points
\beq
q_t, q_t^* = \frac{5}{2} \pm \frac{\sqrt{3} i}{2} = 2 + e^{\pm i\pi/3}
\label{qtriple}
\eeq
which are triple points on the locus ${\cal B}$.  Curves on ${\cal B}$ extend
upwards from $q_t$ and downwards from $q_t^*$ separating region $R_1$ from
$R_3$.  The curves on ${\cal B}$ pass through the origin $u=0$ vertically.
This follows because at $u=0$ there are two $\lambda$'s which are leading and
are degenerate in magnitude.

In region $R_1$, $\lambda_{sq,2,0,1}$ is dominant, so that 
\beq
fl(sq[2 \times \infty,cyc.],q)=fl(sq[2 \times \infty,Mb.],q)=q-2 \quad 
{\rm for} \ \ q \in R_1 \ .
\label{flsqpxy2r1}
\eeq
In region $R_2$, $\lambda_{sq,2,2}$ is dominant, so that 
\beq 
|fl(sq[2 \times \infty,cyc./Mb.],q)|=1 \quad {\rm for} \ \ q \in R_2 \
.
\label{flsqpxy2r2}
\eeq
In region $R_3$, $\lambda_{sq,2,1,1}$ is dominant, so that
\beq
|fl(sq[2 \times \infty,cyc./Mb.],q)|=|q-3| \quad {\rm for} \ \ q \in R_3 \
.
\label{flsqpxy2r3}
\eeq

\subsection{$L_{\lowercase{y}}=3$ Cyclic and M\"obius Strips of the Square 
Lattice}

Our general structural results yield $n_F(3,0)=1$, $n_F(3,1)=3$, $n_F(3,2)=2$,
$n_F(3,3)=1$ with the total number $N_{F,3,\lambda}=7$.  For this $L_y=3$ case
we find that a single term can occur with more than one degree for its
coefficient; in particular, the term $(2-q)$ occurs with both coefficient
$c^{(1)}$ and coefficient $c^{(2)}$.  Given that this phenomenon occurs, the
sum (\ref{nftot}) is not, in general, identical to the sum of distinct
$\lambda$'s but instead is an upper bound on this sum.  This is a type of
behavior that does not occur for the Tutte or chromatic polynomials of strips
of regular planar lattices that we have considered in previous work.  For these
strips, the sums of the corresponding numbers $n_T(L_y,d)$ and $n_P(L_y,d)$
over $d$ for a given value of $L_y$ are equal to the respective numbers of
distinct $\lambda$'s in the Tutte and chromatic polynomials.  From our general
calculation of the Tutte polynomial in \cite{s3a}, we find that the nonzero
terms are
\beq
\lambda_{sq,3,0,1}=q^2-5q+7
\label{flam301}
\eeq
\beq
\lambda_{sq,3,1,1}=2-q
\label{flam311}
\eeq
\beq
\lambda_{sq,3,1,j}=\frac{1}{2}\biggl [ 10-6q+q^2 \pm
(52-56q+28q^2-8q^3+q^4)^{1/2} \biggr ] \quad {\rm for} \ \ j=2,3
\label{flam3123}
\eeq
\beq
\lambda_{sq,3,2,1}=2-q
\label{flam321}
\eeq
\beq
\lambda_{sq,3,2,2}=4-q
\label{flam322}
\eeq
\beq
\lambda_{sq,3,3}=1 \ .
\label{flam33}
\eeq

We  label  the  terms  in  the  flow  polynomial  for  the  M\"obius  strip  as
$\lambda_{sq,L_y,d,\pm,j}$  where   $d,\pm$  means   that  the  term   has  the
coefficient $\pm c^{(d)}$ For the terms $\lambda_{sq,L_y,d,\pm,j}$ occurring in
the flow polynomial for the M\"obius strip with width $L_y=3$ we find
\beq
\lambda_{sq,3,0,+,1}=\lambda_{sq,3,0,1}=q^2-5q+7
\label{flam30p1}
\eeq
\beq
\lambda_{sq,3,0,+,2}=\lambda_{sq,3,2,1}=2-q
\label{flam30p2}
\eeq
\beq
\lambda_{sq,3,0,-,1}=\lambda_{sq,3,2,2}=4-q
\label{flam30m1}
\eeq
\beqs
& & \lambda_{sq,3,1,+,j} = \lambda_{sq,3,1,j+1} \cr\cr 
& = & \frac{1}{2}\biggl [ 10-6q+q^2 \pm
(52-56q+28q^2-8q^3+q^4)^{1/2} \biggr ] \quad {\rm for} \ \ j=1,2
\label{flam31pj}
\eeqs
\beq
\lambda_{sq,3,1,-,1}=\lambda_{sq,3,1,1}=2-q
\label{flam31m1}
\eeq
\beq
\lambda_{sq,3,2,+,1}=\lambda_{sq,3,3}=1 \ .
\label{flam32p1}
\eeq

\bigskip

Then 
\beqs
F(sq[3 \times m,cyc.],q) &=& (\lambda_{sq,3,0,1})^m+
c^{(1)}\sum_{j=1}^3 (\lambda_{sq,3,1,j})^m \cr\cr
& & + c^{(2)}\sum_{j=1}^2 (\lambda_{sq,3,2,j})^m+ c^{(3)}
\label{flowsqpxy3}
\eeqs
and
\beqs
& & F(sq[3 \times m,Mb.],q)=(\lambda_{sq,3,0,+,1})^m+
(\lambda_{sq,3,0,+,2})^m-(\lambda_{sq,3,0,-,1})^m  \cr\cr
& & + c^{(1)}\biggl ( (\lambda_{sq,3,1,+,1})^m+
(\lambda_{sq,3,1,+,2})^m-(\lambda_{sq,3,1,-,1})^m \biggr ) + c^{(2)} \ .
\label{flowsqpxy3mb}
\eeqs

We observe that $F(sq[3 \times m,cyc.],q)$
and $F(sq[3 \times m,Mb.],q)$ always have the factors
$(q-1)(q-2)$.  For $m \ge 3$ odd, $F(sq[3 \times m,cyc.],q)$ also has
the factor $(q-3)$.  The flow numbers for the cyclic $3 \times m$ strip are
therefore the same as for the cyclic $2 \times m$ strip of the square lattice,
given above in eq. (\ref{phisqpxy2}).  Further, we have 
\beq
\phi(sq[3 \times m,Mb.]) = 3 \ .
\label{phisqpxy3mb}
\eeq

We also find 
\beq
F(sq[3 \times m,cyc.],q=3)=\cases{6 & if $m \ge 2$ is even \cr\cr
                                            0 & if $m \ge 1$ is odd}
\label{flowsqpxy3q3}
\eeq
\beq
F(sq[3 \times m,Mb.],q=3)=\cases{2 & if $m \ge 2$ is even \cr\cr
                                             4 & if $m \ge 1$ is odd}
\label{flowsqtpx3q3}
\eeq

\begin{figure}[hbtp]
\centering
\leavevmode
\epsfxsize=4.0in
\begin{center}
\leavevmode
\epsffile{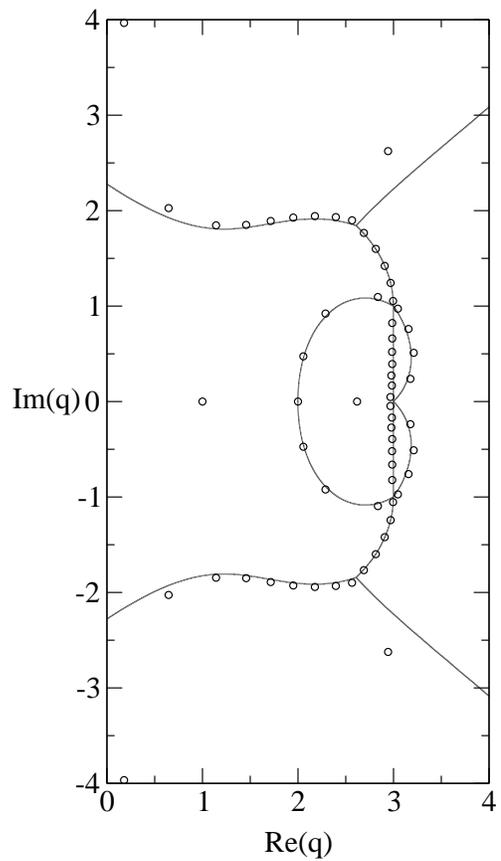}
\end{center}
\caption{\footnotesize{Singular locus ${\cal B}$ in the $q$ plane for $fl(sq,3
\times \infty,cyc./Mb.,q)$ for the $3 \times \infty$ strip of the square
lattice with cyclic or M\"obius (Mb.) boundary conditions. For comparison,
zeros of the flow polynomial $F(sq,3 \times L_x,cyc.,q)$ for $L_x=30$ 
(so that this polynomial has degree equal to $c(G)=61$) are also shown.}}
\label{sqpxy3flow}
\end{figure}
 
\begin{figure}[hbtp]
\centering
\leavevmode
\epsfxsize=4.0in
\begin{center}
\leavevmode
\epsffile{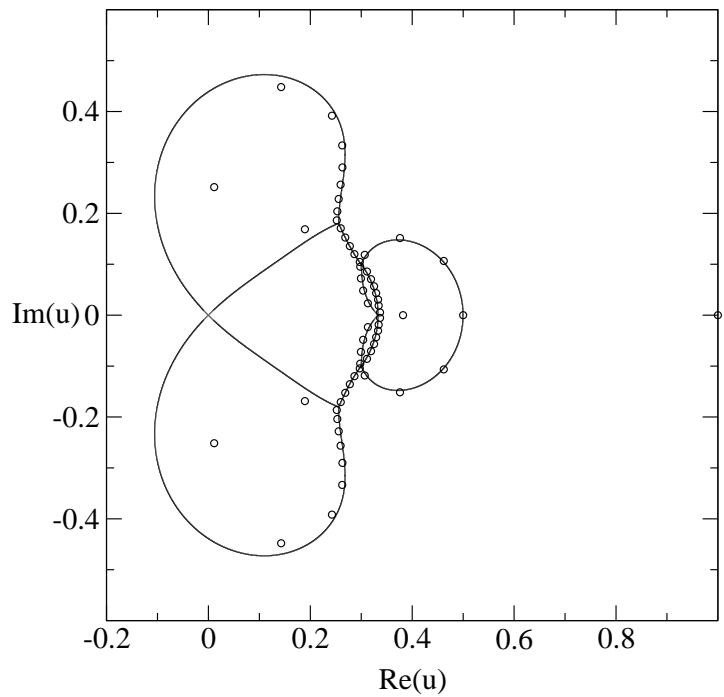}
\end{center}
\caption{\footnotesize{Singular locus ${\cal B}$ in the $u$ plane for $fl(sq,3
\times \infty,cyc./Mb.,q)$ 
for the $3 \times \infty$ strip of the square lattice
with cyclic or M\"obius boundary conditions. For comparison, zeros of the
flow polynomial $F(sq,3 \times L_x,cyc.,q)$ for $L_x=30$ are also shown.}}
\label{sqpxy3uflow}
\end{figure}

The locus ${\cal B}$ for $L_x \to \infty$ is shown in
Fig. \ref{sqpxy3flow}. Again, this locus is noncompact in the $q$ plane.  The
locus is shown in the $u$ plane, where it is compact, in
Fig. \ref{sqpxy3uflow}. The locus separates the $q$ plane into several regions.
Three of these regions, $R_j$, $j=1,2,3$, contain intervals of the real axis,
which are $q \ge 3$ for $R_1$, $2 \le q \le 3$ for $R_2$, and $q < 2$ for
$R_3$.  Hence,
\beq
q_{cf} = 3 \quad {\rm for} \quad \{G\} = sq, \ 3 \times \infty, \ cyc./Mb.
\label{qcsqpxy3}
\eeq
A general feature that we find is that, aside from the trivial case $L_y=1$,
for which ${\cal B}=\emptyset$, ${\cal B}$ crosses the real axis at
$q=2$ and $q=3$ for all of the widths $L_y=2,3,4$ for which we have obtained
exact general formulas for the flow polynomials (see below for the $L_y=4$
case).  For $L_y=3$, the dominant terms in the above-mentioned three regions
are (with an appropriate choice of branch cut for the square root)
$\lambda_{sq,3,1,2}$ in $R_1$, $\lambda_{sq,3,2,2}$ in $R_2$, and
$\lambda_{sq,3,1,2}$ in $R_3$, so that
\beq
fl(sq[3 \times \infty,cyc./Mb.],q)=(\lambda_{sq,3,1,2})^{1/2} \quad
{\rm for} \ \ q \in R_1 \ .
\label{flsqpxy3r1}
\eeq
In region $R_2$, $\lambda_{sq,3,2,2}$ is dominant, so that
\beq 
|fl(sq[3 \times \infty,cyc./Mb.],q)|=|q-4|^{1/2} \quad {\rm for} \ \ q \in
R_2 \ .
\label{flsqpxy3r2}
\eeq
In region $R_3$, $\lambda_{sq,3,1,2}$ is dominant, so that
\beq 
|f(sq[3 \times \infty,cyc./Mb.],q)|=|\lambda_{sq,3,1,2}|^{1/2} \quad {\rm
for} \ \ q \in R_3 \ .
\label{flsqpxy3r3}
\eeq
There is also a pair of small complex-conjugate enclosed regions located to the
upper and lower right of the point $q=3$; in these regions,
$\lambda_{sq,3,2,1}$ is dominant.  The part of ${\cal B}$ forming the boundary
between region $R_2$ and regions $R_4, \ R_4^*$ is the vertical line segment
given by $Re(q)=3$, $-1 \le Im(q) \le 1$.  Four curves on ${\cal B}$ meet at
the intersection points $q=\pm i$, and as one moves further upward above $q=i$,
${\cal B}$ consists of a boundary separating regions $R_1$ and $R_3$.  This
boundary bifurcates at $q \simeq 2.6 + 1.8i$, and as one moves to larger values
of $Im(q)$, one enters the region $R_5$.  Correspondingly, there is the
complex-conjugate region $R_5^*$.   Expanding the equation for the degeneracy
of magnitudes of leading terms $\lambda$ around $u=0$, we have 
\beq 
|1-5u+7u^2|=|1-5u+8u^2+O(u^3)|
\label{eq1mag}
\eeq
Expressing this in terms of polar coordinates, with $u=re^{i\theta}$, and 
expanding around $r=0$, we get the equation, to leading order in $r$, 
\beq
r^2\cos 2\theta = 0 \quad {\rm as} \ \ r \to 0
\label{rteq}
\eeq
The solution to this equation is 
\beq
\theta = \frac{(2j+1)\pi}{4} \ , \quad j=0,1,2,3
\label{thetavals}
\eeq
i.e., $\theta = \pm \pi/4$, $\pm 3 \pi/4$, so that the curves approach $u=0$ 
with these angles.

\bigskip

\subsection{$L_{\lowercase{y}}=4$ Cyclic and M\"obius Strips of the Square
Lattice}

For the flow polynomial of the cyclic $4 \times m$ strip of the square lattice
our general results yield $n_F(4,0)=3$, $n_F(4,1)=6$, $n_F(4,2)=6$, 
$n_F(4,3)=3$, $n_F(4,4)=1$ with the total number $N_{F,4,\lambda}=19$.  
We have calculated the flow polynomial for this case.  We find that the three
$\lambda_{F,4,0,j}$, $j=1,2,3$, with coefficients $c^{(0)}$ are the roots of 
the equation 
\beqs
& & \xi^3-(q^3-8q^2+24q-26)\xi^2 \cr\cr
& &  -(q^5-12q^4+59q^3-149q^2+193q-101)\xi+(q-3)(q-2)^5=0 \ .
\label{eqly4d0}
\eeqs
For the $\lambda$'s with coefficient $c^{(1)}$, we have 
\beq
\lambda_{F,4,1,1}=-(q-2)^2
\label{flam411}
\eeq
\beq
\lambda_{F,4,1,2}=-(q-3)^2 \ .
\label{flam412}
\eeq
The $\lambda_{F,4,1,j}$ for $j=3,4,5,6$ are the roots of the equation
\beqs
& & \xi^4-(q-3)(q^2-6q+12)\xi^3-(q-2)(2q^4-22q^3+96q^2-196q+159)\xi^2 \cr\cr
& & -(q-3)(q^6-16q^5+105q^4-366q^3+717q^2-750q+327)\xi \cr\cr
& & +(q^6-15q^5+93q^4-306q^3+565q^2-555q+227)(q-2)^2=0 \ .
\label{eqly4d1}
\eeqs
For the $\lambda$'s with coefficient $c^{(2)}$, 
\beq
\lambda_{F,4,2,j}=\frac{1}{2}\biggl (-(q-3)^2 \pm 
\sqrt{(q-3)(q^3-5q^2+11q-11)} \ \biggr ) \quad {\rm for} \ \ j=1,2 \ .
\label{flam42j}
\eeq
The $\lambda_{F,4,2,j}$ for $j=3,4,5,6$ are roots of the equation
\beqs
& & \xi^4+(2q^2-13q+22)\xi^3+(q-2)(q^3-13q^2+51q-66)\xi^2 \cr\cr
& & -(2q^5-28q^4+152q^3-402q^2+521q-265)\xi \cr\cr
& & +(q^4-11q^3+43q^2-70q+41)(q-2)^2 = 0 \ .
\label{eqly4d2}
\eeqs
For the $\lambda$'s with coefficient $c^{(3)}$ we find
\beq
\lambda_{F,4,3,1}=q-3
\label{flam431}
\eeq
\beq
\lambda_{F,4,3,2}=q-(3-\sqrt{2})
\label{flam432}
\eeq
\beq
\lambda_{F,4,3,3}=q-(3+\sqrt{2})
\label{flam433}
\eeq
and, in accordance with eq. (\ref{fsqcly}),
%
\beq
\lambda_{F,4,4}=-1 \ .
\label{flam44}
\eeq
Hence 
\beqs
F(sq[4 \times m,cyc.],q) &=& \sum_{j=1}^3 (\lambda_{sq,4,0,j})^m+
c^{(1)}\sum_{j=1}^6 (\lambda_{sq,4,1,j})^m \cr\cr
& & + c^{(2)}\sum_{j=1}^6 (\lambda_{sq,4,2,j})^m + 
      c^{(3)}\sum_{j=1}^3 (\lambda_{sq,4,3,j})^m + c^{(4)}(-1)^m \ . 
\cr\cr
& & 
\label{flowsqpxy4}
\eeqs

For the M\"obius strip of the square lattice with width $L_y=4$ we find
$n_F(4,0,+)=5$, $n_F(4,0,-)=4$, $n_F(4,1,+)=4$, $n_F(4,1,-)=3$,
$n_F(4,2,+)=2$, $n_F(4,2,-)=1$, with 
\beq
\lambda_{sq,4,0,+,j} = \lambda_{sq,4,0,j} \ , \quad j=1,2,3
\label{flammb40plus13}
\eeq
\beq
\lambda_{sq,4,0,+,4} = \lambda_{sq,4,2,1} \ , \quad 
\lambda_{sq,4,0,+,5} = \lambda_{sq,4,2,2}
\label{flammb40plus45}
\eeq
\beq
\lambda_{sq,4,0,-,j} = \lambda_{sq,4,2,j+2} \ , \quad j=1,2,3,4
\label{flammb40minus14}
\eeq
\beq
\lambda_{sq,4,1,+,j} = \lambda_{sq,4,1,j+2} \ , \quad j=1,2,3,4
\label{flammb41plus14}
\eeq
\beq
\lambda_{sq,4,1,-,j} = \lambda_{sq,4,1,j}  \ , \quad j=1,2
\label{flammb41minus12}
\eeq
\beq
\lambda_{sq,4,1,-,3} = \lambda_{sq,4,4} = -1
\label{flammb41minus3}
\eeq
\beq
\lambda_{sq,4,2,+,1} = \lambda_{sq,4,3,2} \ , \quad 
\lambda_{sq,4,2,+,2} = \lambda_{sq,4,3,3} 
\label{flammb42plus12}
\eeq
\beq
\lambda_{sq,4,2,-,1} = \lambda_{sq,4,3,1}
\label{flammb42minus1}
\eeq
so that 
\beqs
& & F(sq[4 \times m,Mb.],q)=\sum_{j=1}^5 (\lambda_{sq,4,0,+,j})^m -
\sum_{j=1}^4 (\lambda_{sq,4,0,-,j})^m \cr\cr 
& & + c^{(1)}\biggl [ \sum_{j=1}^4 (\lambda_{sq,4,1,+,j})^m - 
\sum_{j=1}^3 (\lambda_{sq,4,1,-,j})^m \biggr ] \cr\cr
& & + c^{(2)}\biggl [ \sum_{j=1}^2 (\lambda_{sq,4,2,+,j})^m - 
(\lambda_{sq,4,2,-,1})^m \biggr ] \ .
\label{flowsqpxy4mb}
\eeqs

\bigskip

\begin{figure}[hbtp]
\centering
\leavevmode
\epsfxsize=4.0in
\begin{center}
\leavevmode
\epsffile{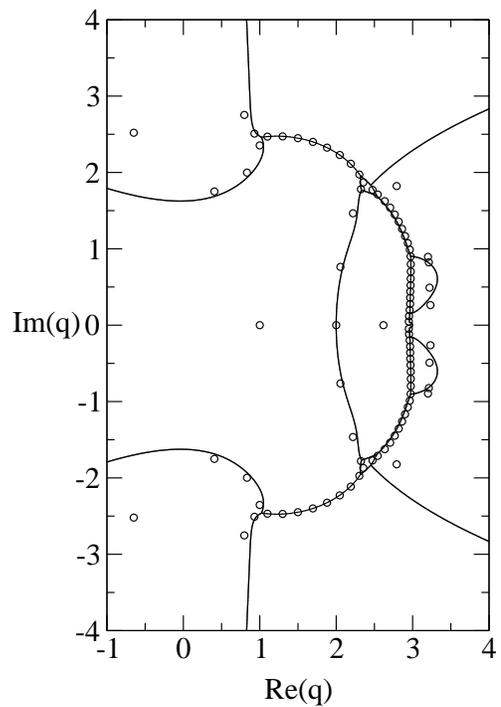}
\end{center}
\caption{\footnotesize{Singular locus ${\cal B}$ in the $q$ plane for $fl(sq,4
\times \infty,cyc./Mb.,q)$ for the $4 \times \infty$ strip of the square
lattice with cyclic or M\"obius boundary conditions. For comparison,
zeros of the flow polynomial $F(sq,4 \times L_x,cyc.,q)$ for $L_x=30$ are 
shown.}}
\label{sqpxy4flow}
\end{figure}

The locus ${\cal B}$ for $L_x \to \infty$ is shown in Fig. \ref{sqpxy4flow}.
This locus is again noncompact in the $q$ plane, containing six curves that
extend infinitely far away from $q=0$. The locus separates the $q$ plane into
several regions.  Three of these regions, $R_j$, $j=1,2,3$, contain intervals
of the real axis, which are $q \ge 3$ for $R_1$, $2 \le q \le 3$ for $R_2$, and
$q < 2$ for $R_3$.  Hence,
\beq
q_{cf} = 3 \quad {\rm for} \quad \{G\} = sq, \ 4 \times \infty, \ cyc./Mb.
\label{qcsqpxy4}
\eeq
The locus ${\cal B}$ again crosses the real axis at $q=2$ and $q=3$.  In region
$R_1$ the dominant term is the root of maximal magnitude of the cubic equation
(\ref{eqly4d0}), which we denote $\lambda_{sq,4,0,jmax}$, so that
\beq
fl(sq[4 \times \infty,cyc./Mb.],q)=(\lambda_{sq,4,0,jmax})^{1/3} \quad
{\rm for} \ \ q \in R_1 \ .
\label{flsqpxy4r1}
\eeq
In region $R_2$, the dominant term is the root of maximal magnitude of the
quartic equation (\ref{eqly4d2}), denoted $\lambda_{sq,4,2,jmax}$, so that
\beq
|fl(sq[4 \times \infty,cyc./Mb.],q)|=|\lambda_{sq,4,2,jmax}|^{1/3} \quad 
{\rm for} \ \ q \in R_2 \ .
\label{flsqpxy4r2}
\eeq
In region $R_3$, the dominant term is the root of maximal magnitude of the
other quartic equation (\ref{eqly4d1}), so that
\beq
|fl(sq[4 \times \infty,cyc./Mb.],q)|=|\lambda_{sq,4,1,jmax}|^{1/3} \quad
{\rm for} \ \ q \in R_3 \ .
\label{flsqpxy4r3}
\eeq
In addition to the regions $R_j$, $j=1,2,3$ that contain intervals of the real
axis, there are also four complex-conjugate pairs of regions away from the real
axis, $R_j$, $R_j^*$, $j=4,5,6,7$.  These can be identified in Fig. 
\ref{sqpxy4flow} as follows: $R_4$ is a small ``bubble'' region containing 
the point $q=3.1 + 0.6i$; $R_6$ contains the point $q=2+3i$ and extends to
complex infinity; $R_7$ contains the point $q=3i$ and extends to complex
infinity; and $R_5$ is a very small region that contains the point
$q=2.4+1.8i$.  

\subsection{On $q_{cf}$ for Cyclic/M\"obius Square-Lattice Strips}

Aside from the trivial case $L_y=1$, all of the cyclic/M\"obius strips of the
square lattice that we have studied have yielded $q_{cf}=3$.  This is the same
value as for the infinite square lattice (see eq. (\ref{qcfsq}).  We are led to
conjecture that $q_{cf}=3$ for cyclic/M\"obius strips of the square lattice for
all $L_y \ge 2$.  Our present finding and conjecture are related to our
previous result that $q_c=3$ for ${\cal B}_W$ for self-dual cyclic strips of
the square lattice \cite{dg,sdg} for all widths considered and our consequent
conjecture that this property $q_c=3$ holds for all cyclic self-dual strips of
the square lattice.  To see this connection, we recall that the dual graph of
the cyclic strip of the square lattice with $L_y \ge 2$ is a cyclic strip of
this lattice with width $L_y-1$ augmented by edges joining all of the vertices
on the upper and lower sides of the strip to two respective external vertices.
The latter strip is rather similar to the cyclic self-dual strips studied in
\cite{dg,sdg}, differing only by the edges joining the lower vertices to the
second external point.  To the extent that these lower edges do not modify
$q_c$, one would then expect that $q_c=3$ for the family of dual graphs, which
is the equivalent to the conjecture that we make here for $q_{cf}$.

Our finding that, aside from the trivial case $L_y=1$, ${\cal B}$ for the $L_y
\times \infty$ cyclic/M\"obius strips of the square lattice crosses the real
axis at $q=2$ for $L_y=2,3,4$ suggests another conjecture, namely that this
property holds for all cyclic or M\"obius strips with width $L_y \ge 2$.  It is
interesting to relate this to our earlier finding that for this class of
cyclic/M\"obius strip graphs of the square lattice, ${\cal B}_W$ crosses the
real axis at $q=0$ and $q=2$ for all widths $1 \le L_y \le 5$ considered
\cite{w,wcy,s4,s5}.  One sees that for the widths $L_y \ge 2$ considered,
${\cal B}_{fl}$ and ${\cal B}_W$ both cross the real axis at $q=2$, while
${\cal B}_W$, but not ${\cal B}_{fl}$, crosses at $q=0$ and ${\cal B}_{fl}$,
but not ${\cal B}_W$, crosses at $q=3$.  For the self-dual strips considered
in \cite{dg,sdg}, for which ${\cal B}_{fl}={\cal B}_W$, there are crossings at
$q=2,3$ but not $q=0$.  The actual behavior of ${\cal B}_{fl}={\cal B}_W$ in 
the $L_y \to \infty$ limit, i.e. for the full square lattice, would presumably
combine these features.

\section{Approach of $fl(sq,L_y \times \infty,q)$ to $fl(sq,q)$} 

It is of interest to use our exact calculations of the flows per face,
$fl(sq,L_y \times \infty,q)$ for the cyclic/M\"obius strips of
the square lattice to study how the values of this function approach the values
for the infinite square lattice as the strip width $L_y$ increases.  This is
done in Table \ref{sqtable} for a range of $q$ values.  For this table, we
define the ratio
\beq
R_{fl}(sq(L_y),FBC_y,q) = \frac{fl(sq,L_y \times \infty,q)}{fl(sq,q)} \ .
\label{rfl}
\eeq
We use calculations of $fl(sq,q)=W(sq,q)$ for the infinite square lattice
obtained using Monte Carlo methods from \cite{ww,w2d}.  Again, we note
that the case $L_y=1$ is atypical since the flow polynomial does not exhibit
exponential growth with $L_x$ and the number of faces is fixed at two,
independent of $L_x=m$, in contrast to all of the strips with $L_y \ge
2$. (Hence, for $L_y=1$ we do not list the $R_{fl}$ values.)  For $L_y \ge 2$
we find that $fl$ approaches the infinite-width value from below and this
approach is not, in general, monotonic. For a given value of $L_y \ge 2$, the
values of $fl$ are closer to those for the infinite lattice for larger $q$.  In
contrast, for cyclic/M\"obius strips and for a given value of $L_y$, $W$
calculated for the infinite strip of width $L_y$ approaches the infinite-width
value from above rather than from below \cite{w2d}.

\begin{table}
\caption{\footnotesize{Comparison of values of $fl(sq(L_y),FBC_y,q)$ with
$fl(sq,q)$ for $3 \le q \le 10$.  For each value of $q$, the quantities in the
upper line are identified at the top and the quantities in the lower line are
the values of $R_{fl}(sq(L_y),FBC_y,q)$. The FBC$_y$ is symbolized as $F$ in
the table.}}
\begin{center}
\begin{tabular}{cccccccc}
\hline 
$q$ & $fl(sq(1),F,q)$ & $fl(sq(2),F,q)$ & $fl(sq(3),F,q)$ & $fl(sq(4),F,q)$ &
$fl(sq,q)$ \\
\hline \\
3 & 1.4142    & 1        & 1        & 1.1740   & 1.53960... \\
  & $-$       & 0.6495   & 0.6495   & 0.7625  \\
4 & 1.73205   & 2        & 1.7989   & 1.9520   & 2.3370(7)  \\
  & $-$       & 0.8558   & 0.7697   & 0.8353  \\
5 & 2         & 3        & 2.7248   & 2.8827   & 3.2510(10) \\
  & $-$       & 0.9228   & 0.8381   & 0.8867  \\
6 & 2.236     & 4        & 3.6802   & 3.8418   & 4.2003(12) \\
  & $-$       & 0.9523   & 0.8762   & 0.9146  \\
7 & 2.449     & 5        & 4.6502   & 4.8138   & 5.1669(15) \\
  & $-$       & 0.9677   & 0.9000   & 0.9317  \\
8 & 2.646     & 6        & 5.6286   & 5.7934   & 6.1431(20) \\
  & $-$       & 0.9767   & 0.9163   & 0.9431  \\
9 & 2.828     & 7        & 6.6124   & 6.7779   & 7.1254(22) \\
  & $-$       & 0.9824   & 0.9280   & 0.9512 \\
10 & 3        & 8        & 7.5998   & 7.7657   & 8.1122(25) \\
  & $-$       & 0.9862   & 0.9368   & 0.9573 \\
\hline
\end{tabular}
\end{center}
\label{sqtable}
\end{table}

\section{Strips of the Square Lattice with Torus and Klein Bottle Boundary 
Conditions} 

It is of interest to study flows on lattice strip graphs that have doubly
periodic boundary conditions.  We carry out this study in the present section
for the square lattice strips, including both the case of torus and Klein
bottle (Kb.) topologies.

\subsection{$L_{\lowercase{y}}=2$ Strips of the Square Lattice with Torus and 
Klein Bottle B.C.} 

This family involves double edges on each transverse slice of the strip. 
Although the chromatic polynomial is insensitive to the presence of multiple
edges, the flow polynomial is sensitive to this property, as is obvious since
in general multiple edges allow more flows satisfying the conservation 
conditions at each vertex. Using our results in \cite{s3a} we have
\beq
\lambda_{sqt,2,0,1}=D_4=q^2-3q+3
\label{flamsqt201}
\eeq
where 
\beq
D_m(q) = \frac{P(C_m,q)}{q(q-1)} = 
\sum_{s=0}^{m-2}(-1)^s {{m-1}\choose {s}} q^{m-2-s}
\label{dk}
\eeq
(where the second equality holds for $m \ge 2$), 
\beq
\lambda_{sqt,2,1,1}=q^2-4q+5
\label{flamsqt211}
\eeq
\beq
\lambda_{sqt,2,2}=1
\label{flamsqt22}
\eeq
and
\beqs
F(sq[2 \times L_x=m,torus],q) & = & (\lambda_{sqt,2,0,1})^m+
c^{(1)}(\lambda_{sqt,2,1,1})^m+c^{(2)} \cr\cr
& = & (q^2-3q+3)^m + (q-1)(q^2-4q+5)^m+(q^2-3q+1) \cr\cr
& & 
\label{fsqpxpy2}
\eeqs
\beqs
F(sq[2 \times L_x=m,Kb.],q) & = & (\lambda_{sqt,2,0,1})^m+
c^{(1)}(\lambda_{sqt,2,1,1})^m-1 \cr\cr
& = & (q^2-3q+3)^m + (q-1)(q^2-4q+5)^m-1 \ . \cr\cr
& & 
\label{fsqtpxpy2}
\eeqs
We have $C_{F,sq,L_y=2,tor}=(q-1)^2$ for the $2 \times m$ torus strip and
$C_{F,sq,L_y=2,Kb.}=q-1$ for the $2 \times m$ Klein bottle strip of the square
lattice. Both flow polynomials have $(q-1)$ as a factor, so that the flow
number $\phi=2$ for these families of graphs.

The locus ${\cal B}$ is noncompact in the $q$ plane and separates this plane
into four regions.  This locus is identical to the locus for ${\cal B}_W$ in
Fig. 5(a) of \cite{wa2}.  The region $R_1$ includes the real segment $q > 2$,
while $R_2$ contain the segment $q < 2$. For this family,
\beq
q_{cf}=2 \quad{\rm for} \quad sq, 2 \times \infty, \ {\rm torus \ or \ Kb.}
\label{qcsqpxpy2}
\eeq
This demonstrates that for a given type of lattice strip and for a given strip
width $L_y$, the imposition of different transverse boundary conditions leads,
in general, to a different locus ${\cal B}$ and, for the case of periodic 
longitudinal boundary conditions where ${\cal B}$ crosses the real axis, can 
lead to different values of $q_{cf}$.  

Above and below $q_{cf}$ there are two complex-conjugate phases that extend up
and down to triple points, from which c.c. curves extend outward to complex
infinity.  In $R_1$, $\lambda_{sqt,2,0,1}$ is dominant, so
\beq
fl(sq[2 \times \infty,torus/Kb.],q)=(q^2-3q+3)^{1/2} \quad {\rm for} \quad
q \in R_1 \ .
\label{flsqtr1}
\eeq
where the notation ``torus/Kb.'' means that the result holds for the strip with
either torus or Klein bottle boundary conditions. 
In region $R_2$, $\lambda_{sqt,2,1,1}$ is dominant, so 
\beq
|fl(sq[2 \times \infty,torus/Kb.],q)|=|q^2-4q+5|^{1/2} \quad {\rm for} 
\quad q \in R_2 \ .
\label{flsqtr2}
\eeq
In regions $R_3$ and $R_3^*$, $\lambda_{sqt,2,2}$ is dominant, so 
\beq
|fl(sq[2 \times \infty,torus/Kb.],q)|=1 \quad {\rm for}
\quad q \in R_3, \ R_3^* \ .
\label{flsqtr3}
\eeq

\subsection{$L_{\lowercase{y}}=3$ Strips of the Square Lattice with Torus and 
Klein Bottle B.C.} 

Here, for the $L_y=3$ strips of the square lattice with torus and Klein 
bottle (Kb.) boundary conditions, which we denote generically as 
$sq3t$ and $sq3kb$, we find 
\beq
N_{F,sq3t,\lambda}=8
\label{nftotsqpxy3}
\eeq
\beq
N_{F,sq3kb,\lambda}=6 \ .
\label{nftotsqtpxy3}
\eeq
These are reductions from the numbers of different terms in the Tutte
polynomials for these strips, which are \cite{s3a} 
$N_{T,sq3t,\lambda}=20$ and $N_{T,sq3kb,\lambda}=12$.  These may be
compared with the numbers for the chromatic polynomials for these strips
\cite{tk} $N_{P,sq3t,\lambda}=8$ and $N_{P,sq3kb,\lambda}=5$.  We have 
\beq
F(sq[3 \times m,torus],q) = \sum_{j=1}^8 c_{sq3t,j} (\lambda_{sq3t,j})^m 
\label{flowpolsqpxpy3}
\eeq
\beq
F(sq[3 \times m,Kb.],q) = \sum_{j=1}^6 c_{sq3kb,j} (\lambda_{sq3kb,j})^m 
\label{flowpolsqtpxpy3}
\eeq
where, in order of increasing degrees of the coefficients (see below) 
\beq
\lambda_{sq3t,1}=q^3-6q^2+14q-13
\label{flam1sq3t}
\eeq
\beq
\lambda_{sq3t,j} = \frac{1}{2}\biggl [ -18+16q-6q^2+q^3 \pm \sqrt{R_{sq3t}} \ 
\biggr ] \quad {\rm for} \ \ j=2,3
\label{flam23sq3t}
\eeq
\beq
R_{sq3t}=256-440q+376q^2-196q^3+64q^4-12q^5+q^6
\label{rsq3t}
\eeq
\beq
\lambda_{sq3t,4}=q-2
\label{flam4sq3t}
\eeq
\beq
\lambda_{sq3t,5}=q-1
\label{flam5sq3t}
\eeq
\beq 
\lambda_{sq3t,6}=q-4
\label{flam6sq3t}
\eeq
\beq
\lambda_{sq3t,7}=q-5
\label{flam7sq3t}
\eeq
\beq
\lambda_{sq3t,8}=-1 \ .
\label{flam8sq3t}
\eeq
In contrast to the situation with cyclic and M\"obius strips, there is not a
1-1 correspondence between terms $\lambda$ and the coefficients that the flow
polynomial inherits as a special case of the Tutte polynomial; here, one of the
terms appears with two different Tutte coefficients.  Specifically, the term
$\lambda_{sq3t,4}$ appears with both the coefficient $2(q-1)$ and $q(q-3)$, so
that its net coefficient is the sum, $(q+1)(q-2)$.  For the coefficients we
thus have
\beq
c_{sq3t,1}=1
\label{csq3t1}
\eeq
\beq
c_{sq3t,2}=c_{sq3t,3}=q-1
\label{csq3t23}
\eeq
\beq
c_{sq3t,4}=(q+1)(q-2)
\label{csq3t4}
\eeq
\beq
c_{sq3t,5}=\frac{1}{2}c_{sq3t,6}=\frac{1}{2}(q-1)(q-2)
\label{csq3t56}
\eeq
\beq
c_{sq3t,7}=\frac{1}{2}q(q-3)
\label{csq3t7}
\eeq
\beq
c_{sq3t,8}=q^3-6q^2+8q-1 \ .
\label{csq3t8}
\eeq
The relation of these with the $c^{(d)}$'s was discussed earlier
\cite{tor4,s5}.  The sum of coefficients is 
$C_{F,sq,L_y=3,tor}=(q-1)^3$ for the $3 \times m$ torus strip.

For the Klein bottle strip we have
\beq
\lambda_{sq3kb,j}=\lambda_{sq3t,j} \ , \quad j=1,2,3
\label{flamsqtpxy3123}
\eeq
\beq
\lambda_{sq3kb,4}=\lambda_{sq3t,5}=q-1
\label{flamsqtpxy34}
\eeq
\beq
\lambda_{sq3kb,5}=\lambda_{sq3t,7}=q-5
\label{flamsqtpxy35}
\eeq
\beq
\lambda_{sq3kb,6}=\lambda_{sq3t,8}=-1 \ .
\label{flamsqtpxy36}
\eeq
Thus the flow polynomial for the $L_y=3$ strip of the square lattice with torus
boundary conditions has two terms, $q-2$ and $q-4$, that are absent from the
flow polynomial of this lattice strip with Klein bottle boundary conditions.
Evidently, the set of terms for the strip with Klein bottle boundary conditions
is a subset of the set of terms for the strip with torus boundary conditions.

The corresponding coefficients are 
\beq
c_{sq3kb,1}=c_{sq3t,1}=1
\label{csq3kb1}
\eeq
\beq
c_{sq3kb,j}=c_{sq3t,j}=q-1 \ , \quad j=2,3
\label{csq3kb23}
\eeq
\beq
c_{sq3kb,4}=c_{sq3t,5}=\frac{1}{2}(q-1)(q-2)
\label{csq3kb4}
\eeq
\beq
c_{sq3kb,5}=c_{sq3t,7}=\frac{1}{2}q(q-3)
\label{csq3kb5}
\eeq
\beq
c_{sq3kb,6}=-(q-1) \ .
\label{csq3kb6}
\eeq
The sum of these coefficients is $C_{F,sq,L_y=3,Kb.}=(q-1)^2$. 
The flow numbers of the $sq,3 \times L_x$ strips with
torus and Klein bottle (kb) boundary conditions are
\beq
\phi(sq[3 \times L_x,torus/Kb.])=2 \ .
\label{phisqpxpy3}
\eeq

The locus ${\cal B}$ is the same for the torus and Klein bottle strips of a
given lattice.  We have proved this type of equality for ${\cal B}_W$ in
\cite{tor4} for chromatic polynomials of torus and Klein bottle strips.  The
proof for the chromatic polynomials uses the construction of a family of
lattice strip graphs which are identical to the torus and Klein bottle strips
for even and odd length $L_x$.  The locus ${\cal B}$ is the accumulation set of
the zeros in the limit $L_x \to \infty$, and this limit can be taken over
either even or odd values of $L_x$, which proves the equality of the loci
${\cal B}$ for the respective $L_x \to \infty$ limits of these two types of
strips.  The same method of proof applies here for the flow polynomials.  Note
that this result requires that all of the dominant terms are the same for the
strips with torus and Klein bottle boundary conditions.  This condition is
satisfied.

\begin{figure}[hbtp]
\centering
\leavevmode
\epsfxsize=4.0in
\begin{center}
\leavevmode
\epsffile{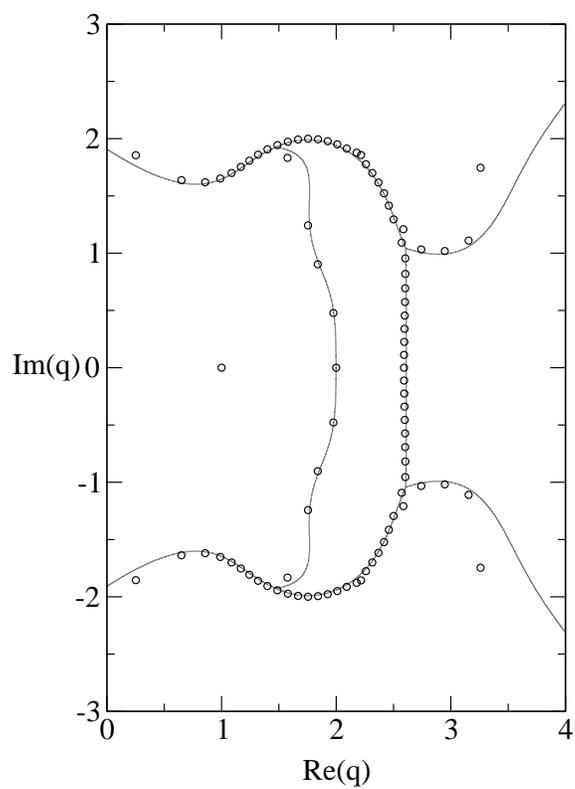}
\end{center}
\caption{\footnotesize{Singular locus ${\cal B}$ in the $q$ plane for $fl(sq,3
\times \infty,torus/Kb.,q)$ for the $3 \times \infty$ strip of the square
lattice with torus or Klein bottle boundary conditions. For comparison,
zeros of the flow polynomial $F(sq,3 \times L_x,torus,q)$ for $L_x=30$ are also
shown.}}
\label{sqpxpy3flow}
\end{figure}
 
\begin{figure}[hbtp]
\centering
\leavevmode
\epsfxsize=4.0in
\begin{center}
\leavevmode
\epsffile{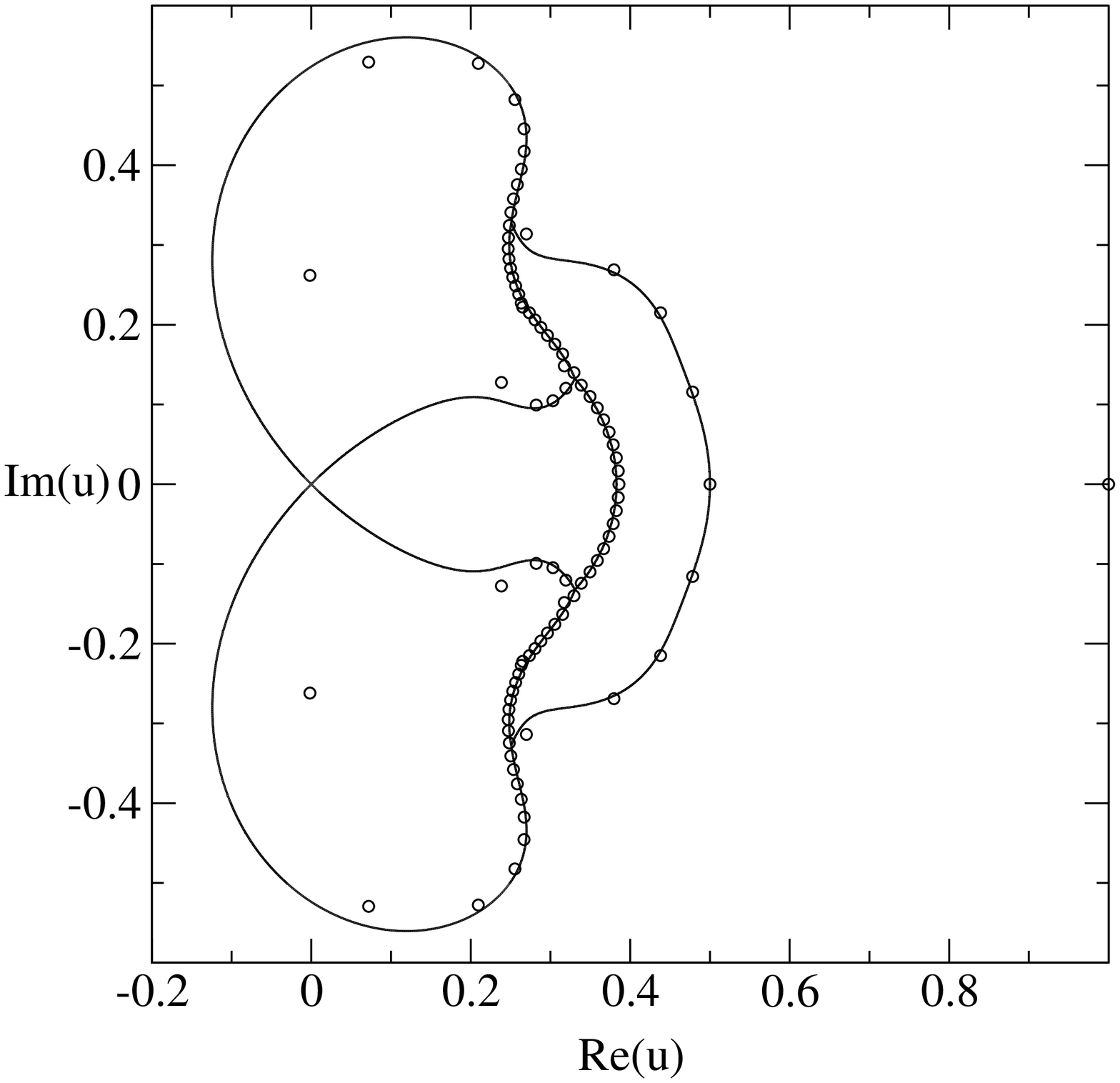}
\end{center}
\caption{\footnotesize{Singular locus ${\cal B}$ in the $u$ plane for $fl(sq,3
\times \infty,torus/Kb.,q)$ for the $3 \times \infty$ strip of the square
lattice with torus or Klein bottle boundary conditions. For comparison,
zeros of the flow polynomial $F(sq,3 \times L_x,torus,q)$ for $L_x=30$ are also
shown.}}
\label{sqpxpy3uflow}
\end{figure}

The locus ${\cal B}$ for the $L_x \to \infty$ limit of the $L_y=3$ square strip
with torus or Klein bottle boundary conditions is shown in the $q$ and $u$
planes in Figs. \ref{sqpxpy3flow} and \ref{sqpxpy3uflow}.  As is evident,
${\cal B}$ is noncompact in the $q$ plane and separates this plane into various
regions.  Three of these regions, $R_j$, $j=1,2,3$, contain intervals of the
real axis, which are $q \ge q_{cf}$ for $R_1$, $2 \le q \le q_{cf}$ for
$R_2$, and $q
< 2$ for $R_3$, where
\beq
q_{cf}=2.6120932... \quad{\rm for} \quad sq, 3 \times \infty, {\rm torus/Kb.}
\label{qcsqpxpy3}
\eeq
This is the larger of the two real roots of the equation
\beq
2q^4-19q^3+71q^2-142q+132=0
\label{qceqly3tor}
\eeq
resulting from the condition of degeneracy in magnitude of the leading terms
$|\lambda_{sq3t,7}|=|\lambda_{sq3t,2}|$.  The fact that $q_{cf}$ increases from
the value 2 for $L_y=2$ to approximately 2.61 for $L_y=3$ is in accord with the
property that as $L_y \to \infty$, $q_{cf} \to 3$ (see eq. (\ref{qcfsq})
below).  The behavior exhibited by the strips of the square lattice with
torus/Klein bottle boundary conditions is thus that $q_{cf}$ progressively
approaches the asymptotic value $q_{cf}=3$ as $L_y$ increases rather than being
equal to this value for each $L_y \ge 2$ as we find for the cyclic/M\"obius
strips of the square lattice. 

The dominant term in region $R_1$ is $\lambda_{sq3t,2}$ so that
\beq
fl(sq[3 \times \infty,torus/Kb.],q)=(\lambda_{sq3t,2})^{1/3} \quad
{\rm for} \quad q \in R_1 \ .
\label{flsqpxpy3r1}
\eeq
This is interesting since, unlike the flow polynomials for lattice strips that
we have studied before, here the dominant term in region $R_1$ does not have
coefficient 1.  We recall that the chromatic polynomials for the recursive
strips that we have considered always have the property that the dominant term
in region $R_1$ (including the positive real $q$ axis) has coefficient 1
\cite{bcc}.  In region $R_2$, $\lambda_{sq3t,7}=q-5$ is dominant, so that
\beq
|fl(sq[3 \times \infty,torus/Kb.],q)|=|q-5|^{1/3} \quad
{\rm for} \quad q \in R_2 \ .
\label{flsqpxpy3r2}
\eeq
In region $R_3$ $\lambda_{sq3t,3}$ is dominant, so
\beq 
|fl(sq[3 \times \infty,torus/Kb.],q)|=|\lambda_{sq3t,3}|^{1/3} \quad
{\rm for} \quad q \in R_3 \ .
\label{flsqpxpy3r3}
\eeq
The boundary on ${\cal B}$ separating regions $R_1$ and $R_2$ is the vertical
line segment in the $q$ plane given by $Re(q)=q_{cf}$ as given above in
eq. (\ref{qcsqpxpy3}) and $-1.04 \le Im(q) \lsim 1.04$. This line segment ends
at the complex-conjugate set of triple points at $q=q_{cf} \pm 1.04i$.  Going
vertically upward and downward away from the real axis along this direction,
one passes into two complex-conjugate regions, $R_4$ and $R_4^*$ in which
$\lambda_{sq3t,1}$ is dominant, so that
\beq
|fl(sq[3 \times \infty,torus/Kb.],q)|=|q^3-6q^2+14q-13|^{1/3} \quad {\rm for} 
\quad \in R_4, \ R_4^* \ .
\label{flsqpxpy3r4}
\eeq
At the point $u=0$, four curves on ${\cal B}$ intersect.  Let us express the 
terms $\lambda_{sq3t,j}$ in terms of the variable $u=1/q$ and defining 
the scaled terms $\lambda_{sq3t,j,u}=u^3\lambda_{sq3t,j}$.  With appropriate
choices of branch cuts for the square root, the equation of degeneracy of
magnitude of leading terms in the vicinity of $u=0$, when expanded in a small
$u$ series, is 
\beq
|1-6u+15u^2 + O(u^3)|=|1-6u+14u^2+O(u^3)| \ .
\label{eqdegu}
\eeq
Using polar coordinates $u=r e^{i\theta}$, this reduces to eq. (\ref{rteq})
as $r \to 0$, which shows that the curves approach $u=0$ with the angles
$\theta = \pm \pi/4$, $\pm 3 \pi/4$.

\section{Self-Dual Strips of the Square Lattice}

In \cite{dg,sdg} we calculated chromatic polynomials and Tutte polynomials for
families of planar self-dual strips of the square lattice with free and cyclic
longitudinal boundary conditions, denoted, respectively, 
$G_D(L_y \times L_x,cyc.)$ and $G_D(L_y \times L_x,free)$.  
Applying the relation (\ref{fpdual}), we infer, in particular, that 
\beq
F(G_D,q) = q^{-1}P(G_D,q)
\label{fpgd}
\eeq
where $G_D$ refers to any of these graphs.  Hence,
\beq
q_{cf} = q_c = 3 \quad {\rm for} \ G_D \ {\rm families} \ .
\label{qcfdg}
\eeq
Similarly, the loci ${\cal B}$ are the same for the $L_x \to \infty$ limits of
the flow and chromatic polynomials for these families of graphs.  A particular
property is that, in contrast to the loci ${\cal B}$ for the other families
studied here, these are compact in the $q$ plane.

\section{Strips of the Square Lattice with Free Boundary Conditions}

It is of interest to compare the flow polynomials for strips with periodic or
twisted periodic longitudinal boundary conditions to those with free boundary
conditions.  These flow polynomials may be calculated directly by iterative
application of the contraction-deletion property or via the chromatic
polynomials of the dual graphs, or as special cases of Tutte polynomials.  We
list some relevant results.  Clearly for a tree graph, the flow polynomial
vanishes.  For strips with free longitudinal boundary conditions, we shall use
a different labelling convention, viz., $m=L_x-1$, than the convention $m=L_x$
used for strips with periodic longitudinal boundary conditions.  For $m \ge 1$
we have
\beq
F(sq[L_y=2,L_x=m+1,free],q)=(q-1)(q-2)^{m-1}
\label{flowsqxy2}
\eeq
\beq
F(sq[L_y=3,L_x=m+1,free],q)=(q-1)(q-2)(q^2-5q+7)^{m-1} \ .
\label{flowsqxy3}
\eeq
Thus, $N_{F,sq,L_y,\lambda}=1$ for $L_y=2,3$.  For both of these families of
strip graphs, the continuous accumulation set of zeros is the empty set since 
the zeros are discrete.

It is convenient to use a generating function to give the results for the cases
$L_y \ge 4$.  This is similar to the generating functions that were utilized
for chromatic polynomials in \cite{strip,strip2} and our subsequent works.  For
the strip of type $G$, length $L_x=m+1$, we write
\beq
\Gamma(G,q,z) = \sum_{m=0}^\infty F(G_m,q)z^m 
\label{gammazcyl}
\eeq
where 
\beq
\Gamma(G,q,z) = \frac{ {\cal N}(G,q,z)}{{\cal D}(G,q,z)}
\label{gammazcalccyl}
\eeq
with
\beq
{\cal N}(G,q,z) = \sum_{j=0}^{d_{\cal N}} A_{G,j}(q) z^j
\label{n}
\eeq
and
\beq
{\cal D}(G,q,z) = 1 + \sum_{j=1}^{d_{\cal D}} b_{G,j}(q) z^j
\label{d}
\eeq
\beq
{\cal N}(sq[L_y=4,free],q,z) = (q-1)(q-2)z[(q-2) - (q-1)(q-3)z]
\label{sqfreely4n}
\eeq
\beqs
& & {\cal D}(sq[L_y=4,free],q,z) = 1 - (q^3-8q^2+24q-26)z \cr\cr
& & - (q^5-12q^4+59q^3-149q^2+193q-101)z^2+(q-2)^5(q-3)z^3 \ .
\label{sqfreely4d}
\eeqs
Thus, $N_{F,sq,L_y=4,free,\lambda}=3$.  The locus ${\cal B}$ for the $L_x \to
\infty$ limit is compact and consists of complex-conjugate pairs of arcs
together with a self-conjugate arc that crosses the real axis at $q \simeq
2.6252$.  This is similar to what we found for the loci ${\cal B}_W$ for strips
with free boundary conditions in earlier work \cite{strip,strip2}.  Although
${\cal B}$ crosses the real axis for this strip, it is not guaranteed to cross
the real axis when one uses free longitudinal boundary conditions, as will be
illustrated by an explicit example below.

We have also calculated the generating function for
$F(sq[L_y=5,L_x,free],q)$
and find that $N_{F,sq,L_y=5,free,\lambda}=4$, with 
\beq
A_{sq,5,free,0} = 0
\label{sqfreely5a0}
\eeq
\beq
A_{sq,5,free,1} = (q-1)(q-2)^3
\label{sqfreely5a1}
\eeq
\beq
A_{sq,5,free,2} = -(q-1)(q-2)(q^5-8q^4+22q^3-20q^2-5q+9)
\label{sqfreely5a2}
\eeq
\beqs
A_{sq,5,free,3} & = &
(q-1)(q-2)(2q^7-29q^6+178q^5-598q^4+1186q^3-1387q^2 \cr\cr
& & +884q-237)
\label{sqfreely5a3}
\eeqs
\beq
A_{sq,5,free,4} = -(q-1)(q-2)^4(q^6-14q^5+79q^4-229q^3+359q^2-288q+91)
\label{sqfreely5a4}
\eeq
\beq
b_{sq,5,free,1} = -q^4+10q^3-44q^2+97q-88   
\label{sqfreely5b1}
\eeq
\beq
b_{sq,5,free,2} = -q^7+19q^6-156q^5+726q^4-2085q^3+3711q^2-3790q+1708
\label{sqfreely5b2}
\eeq
\beqs
b_{sq,5,free,3} & = & 
(q-3)(2q^8-39q^7+337q^6-1685q^5+5335q^4-10959q^3+14264q^2 \cr\cr
& & -10753q+3595)
\label{sqfreely5b3}
\eeqs
\beqs
b_{sq,5,free,4} & = &
-(q-2)^2(q^9-23q^8+235q^7-1401q^6+5376q^5-13785q^4+23647q^3 \cr\cr
& & -26198q^2+17033q-4964) \ .
\label{sqfreely5b4}
\eeqs
The locus ${\cal B}$ is comprised of arcs and is similar to, although more
complicated than, that for the $L_y=4$ strip. 

\section{Strips of the Square Lattice with Cylindrical Boundary Conditions}

For strips of the square lattice with cylindrical boundary conditions we
calculate, for $m \ge 1$, 
\beqs
F(sq[L_y=2,L_x=m+1,cyl.],q) & = & (q-1)(q-2)^2 (D_4)^{m-1} \cr\cr 
& = & (q-1)(q-2)^2(q^2-3q+3)^{m-1}
\label{flowsqxpy2}
\eeqs
\beq
F(sq[L_y=3,L_x=m+1,cyl.],q)=(q-1)(q-2)(q-3)^2(q^3-6q^2+14q-13)^{m-1} \ .
\label{flowsqxpy3}
\eeq
For both of these families of strip graphs, the continuous accumulation set of
zeros is empty since the zeros are discrete.

The coefficients of the generating function for $F(sq[L_y=4,L_x,cyl.],q)$ are
\beq
A_{sq,4,cyl,0} = q-1
\label{sqcylly4a0}
\eeq
\beq
A_{sq,4,cyl,1} = -(q-1)(3q^3-18q^2+35q-18)
\label{sqcylly4a1}
\eeq
\beq
A_{sq,4,cyl,2} = (q-1)^2(q-3)(q^3-7q^2+13q-9)
\label{sqcylly4a2}
\eeq
\beq
b_{sq,4,cyl,1} = -q^4+8q^3-29q^2+55q-46
\label{sqcylly4b1}
\eeq
\beq
b_{sq,4,cyl,2} = q^6-12q^5+61q^4-169q^3+269q^2-231q+85 \ . 
\label{sqcylly4b2}
\eeq
Thus, $N_{F,sq,L_y=4,cyl,\lambda}=2$, and the terms are 
\beq
\lambda_{F,sq,L_y=4,cyl,j}=\frac{1}{2}\biggl [ q^4-8q^3+29q^2-55q+46 \pm 
\sqrt{R_{sq4c}} \ \biggr ]
\label{flamsq4cyl}
\eeq
where
\beq
R_{sq4c}=q^8-16q^7+118q^6-526q^5+1569q^4-3250q^3+4617q^2-4136q+1776 \ .
\label{rsq4c}
\eeq
The denominator ${\cal D }$ of the generating function is the same as that for
the generating function for chromatic polynomials of the $4 \times L_x$ strip
of the square lattice with cylindrical boundary conditions, and consequently,
in the $L_x \to \infty$ limit, the locus ${\cal B}$ is also the same as the
locus ${\cal B}_W$ for the $4 \times \infty$ square-lattice strip with
cylindrical boundary conditions, shown as Fig. 3(a) in \cite{strip2}.  This
locus is compact and consists of a complex-conjugate pair of arcs together with
a self-conjugate arc that crosses the real axis (at $q \simeq 2.30$) and a very
short line segment emanating outward from this crossing point on the real axis.
The value of $q_{cf}$, which is given by the right-hand end of the very short
line segment, occurs at $q \simeq 2.35$.  The locus does not separate the $q$
plane into different regions.
   
The coefficients of the generating function for
$F(sq[L_y=5,L_x,cyl.],q)$ are
\beq
A_{sq,5,cyl,0} = q-1
\label{sqcylly5a0}
\eeq
\beq
A_{sq,5,cyl,1} = -(q-1)(4q^4-35q^3+115q^2-157q+62)
\label{sqcylly5a1}
\eeq
\beq
A_{sq,5,cyl,2} = -(q-1)^2(q^6-7q^5-q^4+133q^3-458q^2+645q-348)
\label{sqcylly5a2}
\eeq
\beq
b_{sq,5,cyl,1} = -q^5+10q^4-46q^3+124q^2-198q+148
\label{sqcylly5b1}
\eeq
\beq
b_{sq,5,cyl,2} =
q^8-19q^7+159q^6-767q^5+2339q^4-4627q^3+5800q^2-4212q+1362 \ .
\label{sqcylly5b2}
\eeq 
Thus, $N_{sq,L_y=5,cyl.,\lambda}=2$.  As was the case with $L_y=4$, the
denominator ${\cal D }$ of the generating function is the same as that for the
generating function for chromatic polynomials of the $5 \times L_x$ strip of
the square lattice with cylindrical boundary conditions, and consequently, in
the $L_x \to \infty$ limit, the locus ${\cal B}$ is also the same as the locus
${\cal B}_W$ for the $5 \times \infty$ square-lattice strip with cylindrical
boundary conditions, shown as Fig. 2 of \cite{s4}. As before, ${\cal B}$ is
compact and consists of five arcs, four of which form two complex-conjugate
pairs and one of which is self-conjugate.  The self-conjugate arc crosses the
real axis at $q_{cf} \simeq 2.69168$.  The endpoints of the arcs occur at the
ten zeros of the polynomial $b_{sq,5,cyl,1}^2-4b_{sq,5,cyl,2}$, which square
roots that occur in $\lambda_{sq,L_y=5,cyl,j}$ have branch point singularities.
The arcs comprising ${\cal B}$ do not separate the $q$ plane into different
regions.

\section{$L_{\lowercase{y}}=2,3$ Cyclic Strip of the Honeycomb Lattice}

The $L_y=2$ cyclic or M\"obius strip of the honeycomb lattice can be
constructed by starting with the $L_y=2$ cyclic or M\"obius strip of the square
lattice and inserting degree-2 vertices on each horizontal (longitudinal) edge.
Because the insertion of degree-2 vertices does not affect the flow polynomial,
it follows that, for $BC_x=FBC_x,PBC_x,TPBC_x$ and $L_y=2$, we have 
\beq
F(hc[2 \times L_x,FBC_y,BC_x],q) = F(sq[2 \times L_x,FBC_y,BC_x],q)
\label{fhcfsqly2}
\eeq
and
\beq
fl(hc[2 \times \infty,cyc./Mb],q)=fl(sq[2 \times \infty,cyc./Mb],q) 
\label{flhcflsqly2}
\eeq
where the expressions for $F(sq,2 \times L_x,FBC_y,BC_x,q)$ with the various
longitudinal boundary conditions were given above.  

For $L_y=3$ we calculate 
\beqs
F(hc[3 \times m,cyc.],q) & = & (\lambda_{hc,3,0,1})^m + 
c^{(1)}\sum_{j=1}^3 (\lambda_{hc,3,1,j})^m \cr\cr
& + & c^{(2)}\sum_{j=1}^2 (\lambda_{hc,3,2,j})^m + c^{(3)}
\label{flowhcpxy3}
\eeqs
where
\beq
\lambda_{hc,3,0,1}=(q-3)^2
\label{flamhc301}
\eeq
\beq
\lambda_{hc,3,2,j}= \frac{1}{2}\biggl [ 7-2q \pm \sqrt{13-4q} \ \biggr ]
 \quad {\rm for} \quad j=1,2 
\label{flamhc32j}
\eeq
\beq
\lambda_{hc,3,3}=1
\label{flamhc33}
\eeq
and the $\lambda_{hc,3,1,j}$, $j=1,2,3$ are the roots of the equation 
\beq
\xi^3-(q-3)(q-5)\xi^2-(q-3)^2(2q-5)\xi-(q-2)^2(q-3)^2=0 \ .
\label{eqflam31j}
\eeq
Hence, 
\beq
\phi(hc[3 \times m,cyc.]) = 4 \ .
\label{phihcpxy3cyc}
\eeq

\begin{figure}[hbtp]
\centering
\leavevmode
\epsfxsize=4.0in
\begin{center}
\leavevmode
\epsffile{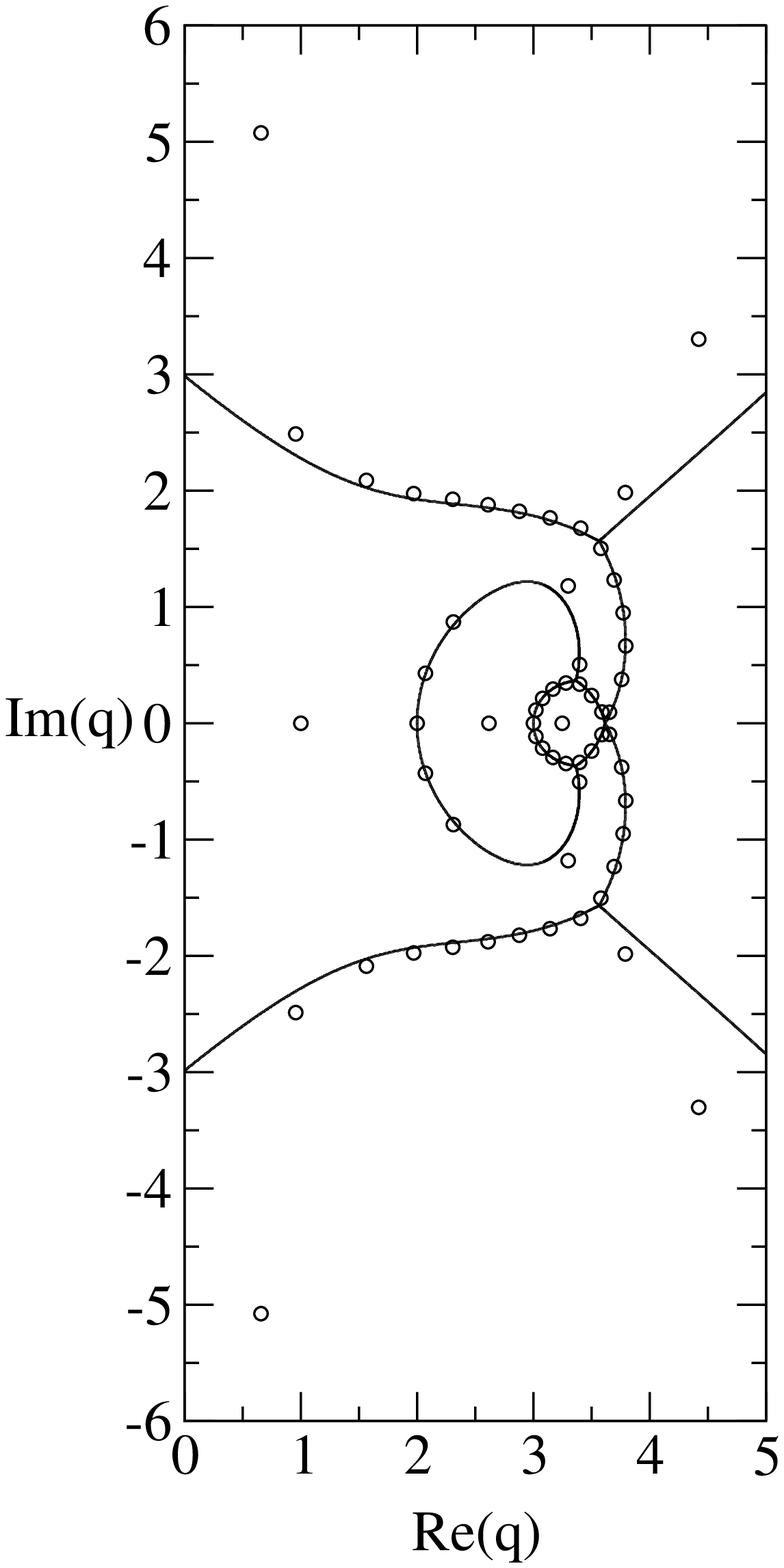}
\end{center}
\caption{\footnotesize{Singular locus ${\cal B}$ in the $q$ plane for $fl(hc,3
\times \infty,cyc./Mb,q)$ for the $3 \times \infty$ strip of the honeycomb
lattice with cyclic or M\"obius boundary conditions. For comparison, zeros of
the flow polynomial $F(hc,3 \times L_x,cyc.,q)$ for $L_x=30$ are also shown.}}
\label{hpxy3flow}
\end{figure}
 
\begin{figure}[hbtp]
\centering
\leavevmode
\epsfxsize=4.0in
\begin{center}
\leavevmode
\epsffile{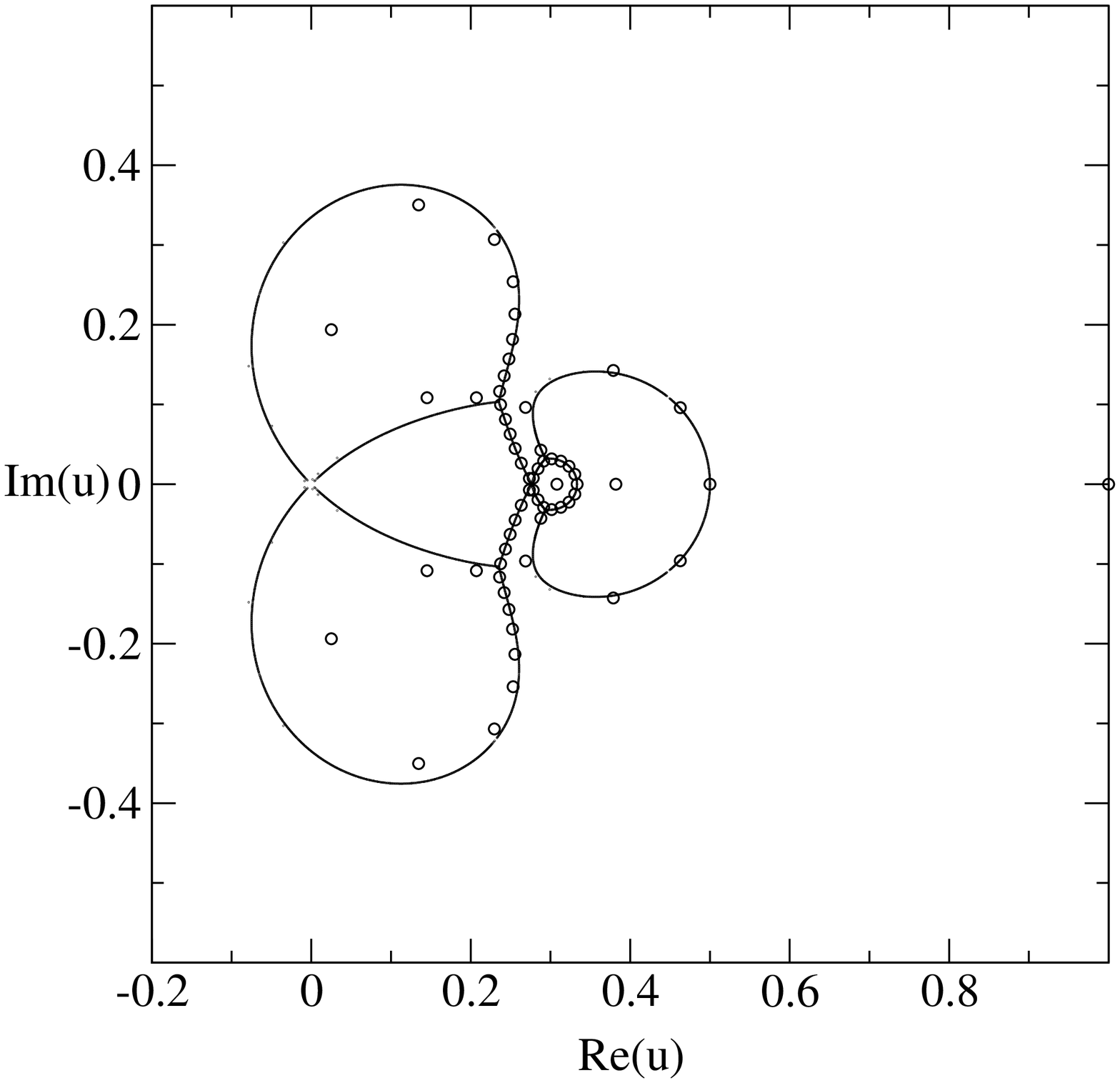}
\end{center}
\caption{\footnotesize{Singular locus ${\cal B}$ in the $u$ plane for $fl(hc,3
\times \infty,cyc./Mb,q)$ for the $3 \times \infty$ strip of the honeycomb
lattice with cyclic or M\"obius boundary conditions. For comparison,
zeros of the flow polynomial $F(hc,3 \times L_x,cyc.,q)$ for $L_x=30$, 
expressed in terms of $u$, are also shown.}}
\label{hpxy3uflow}
\end{figure}

The locus ${\cal B}$ for the $L_x \to \infty$ limit of the cyclic or M\"obius
$L_y=3$ strips of the honeycomb lattice is shown in the $q$ plane in Fig.
\ref{hpxy3flow} and in the $u$ plane in Fig. \ref{hpxy3uflow}. We have
\beq
q_{cf} = \frac{5+\sqrt{5}}{2} = 3.61803... \quad {\rm for} \quad 
hc, 3 \times \infty,cyc./Mb.
\label{qchcpxy3}
\eeq
This is already within 10 \% of the asymptotic value $q_{cf}=4$ for the 2D
honeycomb lattice (see eq. (\ref{qcfhc}) below).  The locus ${\cal B}$ is again
noncompact in the $q$ plane, passing through the origin of the $u$ plane.  The
locus separates the $q$ plane into several regions, which contain intervals of
the real axis: $R_j$, $j=1,2,3,4$, which include the respective intervals (1)
$q \ge q_{cf}$, (2) $3 \le q \le q_{cf}$, (3) $2 \le q \le 3$, and (4) $q \le
2$.  In region $R_1$, the dominant term is the root of the cubic equation
(\ref{eqflam31j}) with maximal magnitude, which we denote as
$\lambda_{hc,3,1,1}$, so that
\beq
fl(hc[3 \times \infty,cyc./Mb],q)=(\lambda_{hc,3,1,1})^{1/2} \quad {\rm
for} \quad q \in R_1 \ .
\label{fhcpxy3r1}
\eeq
In region $R_2$, $\lambda_{hc,3,3}=1$ is dominant, so
\beq
|fl(hc[3 \times \infty,cyc./Mb],q)|=1 \quad {\rm for} \quad q \in R_2 \
.
\label{fhcpxy3r2}
\eeq
In region $R_3$, $\lambda_{hc,3,2,1}$ is dominant, so 
\beq
|fl(hc[3 \times \infty,cyc./Mb],q)|=\left | \frac{1}{2}(7-2q+\sqrt{13-4q} \ )
\right |^{1/2} \quad {\rm for} \quad q \in R_3 \ .
\label{fhcpxy3r3}
\eeq
In region $R_4$ the maximal root of the cubic equation (\ref{eqflam31j}) is
dominant, with a result analogous to (\ref{fhcpxy3r1}) for $fl$.  In regions 
$R_5$ and $R_5^*$ the dominant $\lambda$ is $\lambda_{hc,3,0,1}$, so that
\beq
|fl(hc[3 \times \infty,cyc./Mb],q)|=|q-3| \quad {\rm for} \quad q \in 
R_5, \ R_5^* \ .
\label{fhcpxy3r5}
\eeq
The curves cross the origin of the $u$ plane with the angles $\pm \pi/4$ and
$3\pi/4$.

\section{$L_{\lowercase{y}}=2$ Cyclic Strip of the Triangular Lattice} 

Next, we consider strips of the triangular lattice.  For the $L_y=2$ cyclic 
strip we calculate 
\beq
\lambda_{tri,2,0,1}=(q-2)^2
\label{flamtpxy2_201}
\eeq
\beq
\lambda_{tri,2,1,j}=\frac{1}{2}\biggl [ 6-4q+q^2 \pm (q-2)\sqrt{8-4q+q^2} \ 
\biggr ]
\label{flamtpxy2_21j}
\eeq
where $j=1,2$ for the $\pm$ sign before the square root, and 
\beq
\lambda_{tri,2,2}=1 \ .
\label{flamtpxy2_22}
\eeq
Then
\beq
F(tri[2 \times m,cyc.],q)=(q-2)^{2m}+
c^{(1)}\Bigl [ (\lambda_{tri,2,1,1})^m + (\lambda_{tri,2,1,1})^m \Bigr ] + 
c^{(2)} \ . 
\label{flowtripxy2}
\eeq
Thus, $N_{F,tri,L_y=2,cyc,\lambda}=4$, and for the cyclic strip of the
triangular lattice of width $L_y=2$,
\beq
n_F(tri,2,0)=1, \quad n_F(tri,2,1)=2, \quad n_F(tri,2,2)=1
\label{nftripxy2d}
\eeq
with $n_F(tri,2,d)=0$ for $d \ge 3$.  For this strip 
the sum of coefficients is
\beq
C_{F,tri,L_y=2} = q(q-1) \quad {\rm for} \quad tri, \ 2 \times L_x, cyc. 
\label{csumtripxy2cyc}
\eeq

Using our exact solution, we observe that $F(tri[2 \times m,cyc.],q)$
has only the common factor $(q-1)$ and hence 
\beq
\phi(tri[2 \times m,cyc.]) = 2 \ .
\label{phitripxy2}
\eeq
Further, we find that 
\beq
\phi(tri[2 \times m,Mb.]) = 3 \ .
\label{phitripxy2mb}
\eeq

\begin{figure}[hbtp]
\centering
\leavevmode
\epsfxsize=4.0in
\begin{center}
\leavevmode
\epsffile{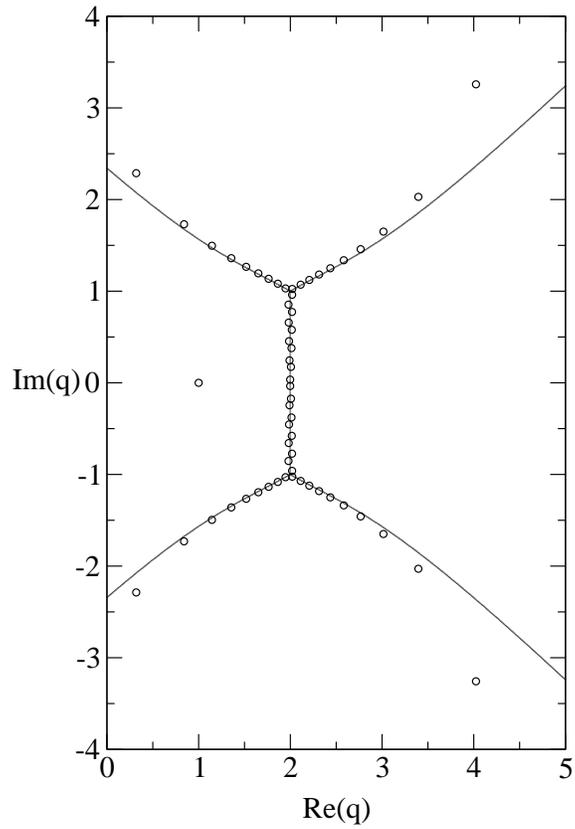}
\end{center}
\caption{\footnotesize{Singular locus ${\cal B}$ in the $q$ plane for $fl(tri,2
\times \infty,cyc./Mb,q)$ for the $2 \times \infty$ strip of the triangular 
lattice with cyclic or M\"obius boundary conditions. For comparison,
zeros of the flow polynomial $F(tri,2 \times L_x,cyc.,q)$ for $L_x=30$ (so that
this polynomial has degree 61) are also shown.}}
\label{tpxy2flow}
\end{figure}
 
\begin{figure}[hbtp]
\centering
\leavevmode
\epsfxsize=4.0in
\begin{center}
\leavevmode
\epsffile{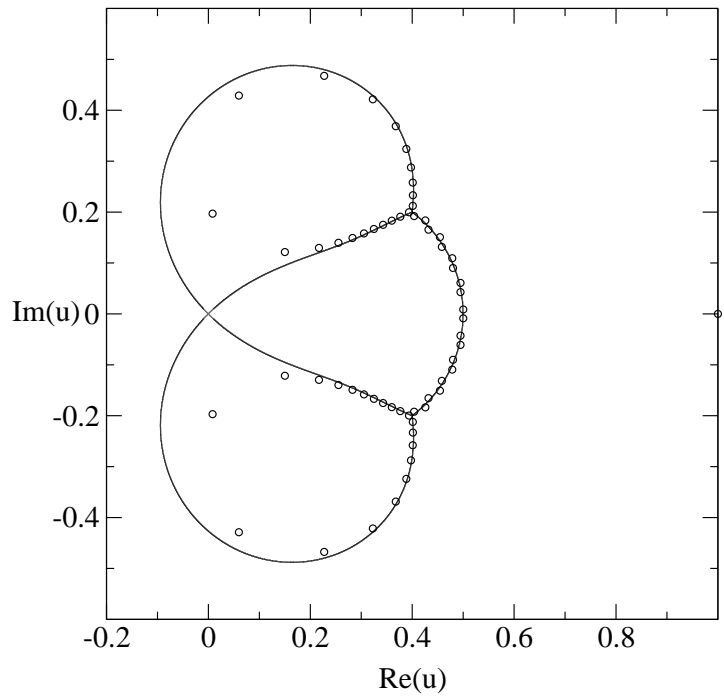}
\end{center}
\caption{\footnotesize{Singular locus ${\cal B}$ in the $u$ plane for $fl(tri,2
\times \infty,cyc./Mb,q)$ for the $2 \times \infty$ strip of the triangular
lattice with cyclic or M\"obius boundary conditions. For comparison,
zeros of the flow polynomial $F(tri,2 \times L_x,cyc.,q)$ for $L_x=30$,
expressed in terms of $u$, are also shown.}}
\label{tpxy2uflow}
\end{figure}

The locus ${\cal B}$ for the cyclic and M\"obius $L_y=2$ strips of the
triangular lattice is shown in the $q$ plane in Fig. \ref{tpxy2flow} and in the
$u$ plane in Fig. \ref{tpxy2uflow}.  This locus is noncompact in the $q$ plane
and divides this plane into four regions, $R_1$, $R_2$, and the
complex-conjugate regions $R_3$ and $R_3^*$.  Regions $R_1$ and $R_2$ contain
the respective real intervals $q > 2$ and $q < 2$ and are contiguous along a
vertical line segment extending from $q=2+i$ to $2-i$.  Triple points on ${\cal
B}$ occur at these endpoints $q=2\pm i$. Evidently,
\beq
q_{cf} = 2 \quad {\rm for} \quad tri, 2 \times \infty,cyc./Mb. \ . 
\label{qctripxy2}
\eeq
Along the vertical line segment from $2+i$ to $2-i$,
$|\lambda_{tri,2,1,1}|=|\lambda_{tri,2,1,2}|=\lambda_{tri,2,2}=1$.  Regions
$R_3$ and $R_3^*$ extend upward and downward, respectively, from the
complex-conjugate triple points at $q=2+i$ and $q=2-i$.  In region $R_1$,
$\lambda_{tri,2,1,1}$ is dominant so that
\beq
fl(tri[2 \times \infty,cyc.],q)=f(tri[2 \times \infty,Mb.],q)=
(\lambda_{tri,2,1,1})^{1/2} \quad {\rm for} \quad q \in R_1 \ .
\label{ftripxy2region1}
\eeq
In region $R_2$, $\lambda_{tri,2,1,2}$ is dominant, so that 
\beq
|fl(tri[2 \times \infty,cyc./Mb.],q)|=|\lambda_{tri,2,1,2}|^{1/2} \quad 
{\rm for} \quad q \in R_2 \ .
\label{ftripxy2region2}
\eeq
In regions $R_3$ and $R_3^*$, $\lambda_{tri,2,0,1}$ is dominant, so that
\beq
|fl(tri[2 \times \infty,cyc./Mb.],q)|=|q-2| \quad {\rm for} \quad q \in 
R_3, \ R_3^* \ .
\label{ftripxy2region3}
\eeq
Some special values include $fl(2)=1$, $|fl(1)|=[(3+\sqrt{5})/2]^{1/2}$, and 
$|fl(0)|=[3+2\sqrt{2}]^{1/2}$. 

To show that ${\cal B}$ passes through $u=0$, we calculate the corresponding
scaled terms $\lambda_{tri,2,d,j,u} = \lambda_{tri,2,d,j}/q^2$ with $u=1/q$ and
see that the degeneracy condition of dominant $\lambda_{tri,2,d,j,u}$'s has a
solution at $u=0$.  The dominant $\lambda_{tri,2,d,j,u}$'s in the vicinity of
$u=0$ are
\beq
\lambda_{tri,2,0,1,u}=(1-2u)^2
\label{flam201tripxy2u}
\eeq
\beq
\lambda_{tri,2,1,1,u}=\frac{1}{2}\biggl [1-4u+6u^2+(1-2u)\sqrt{1-4u+8u^2} \ 
\biggr ] \ .
\label{flam211tripxy2u}
\eeq
For $u \to 0$, the latter term has the Taylor series expansion 
\beq
\lambda_{tri,2,1,1,u}=1-4u+6u^2-u^4+O(u^5) \ .
\label{flam211utaylor}
\eeq
Thus, in the same way as before, introducing polar coordinates and expanding
the equation of the degeneracy of magnitudes of leading terms in the vicinity 
of $u=0$ yields the condition $r^2 \cos 2\theta = 0$, thereby showing that 
four branches of ${\cal B}$ approach the origin of the $u$ plane at the angles
given by (\ref{thetavals}), i.e., $\theta=\pm \pi/4$ and $\theta=\pm 3\pi/4$.

\section{$L_{\lowercase{y}}=3$ Cyclic Strip of the Triangular Lattice} 

For the $L_y=3$ cyclic strip of the triangular lattice we 
find the following results: 
\beqs
& & \lambda_{tri,3,0,j}=\frac{1}{2}\biggl [q^4-7q^3+21q^2-33q+23 \pm \cr\cr
& & (q^8-14q^7+91q^6-360q^5+949q^4-1708q^3+2047q^2-1486q+497)^{1/2} \ 
\biggr ] \, \quad j=1,2 \cr\cr
& &
\label{flamtpxy3_0j}
\eeqs
\beq
\lambda_{tri,3,1,j}=q^2-4q+5 \ .
\label{flamtpxy3_11}
\eeq
The $\lambda_{tri,3,1,j}$ for $j=2,3,4$ are the roots of the equation 
\beqs
& & \xi^3-(q^4-7q^3+22q^2-37q+30)\xi^2 \cr\cr
& &
+(q^6-11q^5+52q^4-134q^3+200q^2-168q+69)\xi-(2q^4-15q^3+42q^2-51q+24)=0
\cr\cr
& & 
\label{trieq33}
\eeqs
\beq
\lambda_{tri,3,2,1}=q^2-3q+3
\label{flamtri321}
\eeq
\beq
\lambda_{tri,3,2,j}=\frac{1}{2}\biggl [ q^2-5q+9 \pm 
(q^4-10q^3+43q^2-90q+73)^{1/2} \ \biggr ] \ , \quad j=2,3
\label{flamtri32j}
\eeq
\beq
\lambda_{tri,3,3}=1 \ .
\label{flamtri33}
\eeq
Hence, $N_{F,tri,L_y=3,cyc.,\lambda}=10$, $n_F(tri,3,0)=2$, 
$n_F(tri,3,1)=4$, $n_F(tri,3,2)=3$, $n_F(tri,3,3)=1$, and 
\beqs
& & F(tri[3 \times m,cyc.],q)=\sum_{j=1}^2 (\lambda_{tri,3,0,j})^m+
c^{(1)}\sum_{j=1}^4 (\lambda_{tri,3,1,j})^m + \cr\cr
& & c^{(2)}\sum_{j=1}^3 (\lambda_{tri,3,2,j})^m+ c^{(3)} \ .
\label{flowtripxy3}
\eeqs

Using our exact result (\ref{phitripxy3}) we observe that 
$F(tri[3 \times m,cyc.],q)$ has only the common factor $(q-1)$, so that
\beq
\phi(tri[3 \times m,cyc.]) = 2 \ .
\label{phitripxy3}
\eeq
This can also be derived as a corollary of the basic theorem discussed in
section (\ref{qexistence}), that a bridgeless graph admits a 2-flow if and only
if all of its vertex degrees are even; here all of the vertices of the 
$tri[3 \times m,cyc.]$ are even.

For the cyclic strip of the triangular lattice of width
$L_y=3$, the sum of coefficients is
\beq
C_{F,tri,L_y=3} = q(q-1)^2 \quad {\rm for} \quad tri, \ 3 \times L_x, cyc. 
\label{csumtripxy3cyc}
\eeq

\begin{figure}[hbtp]
\centering
\leavevmode
\epsfxsize=4.0in
\begin{center}
\leavevmode
\epsffile{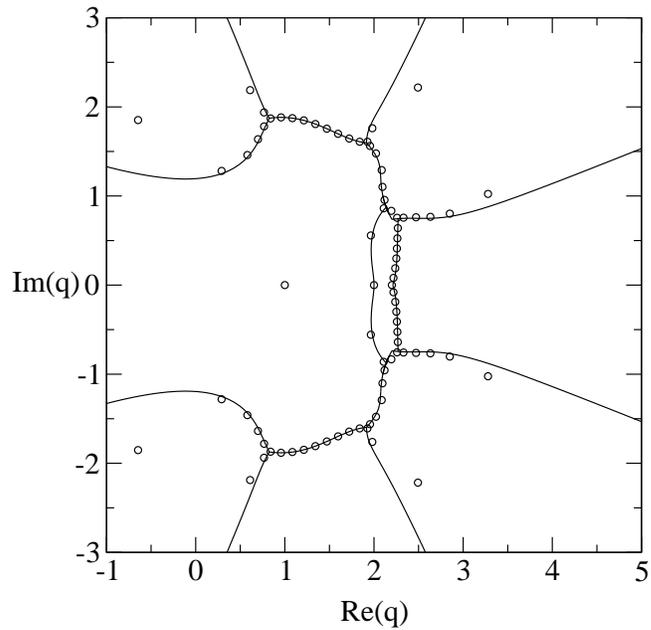}
\end{center}
\caption{\footnotesize{Singular locus ${\cal B}$ in the $q$ plane for $fl(tri,3
\times \infty,cyc./Mb,q)$ for the $3 \times \infty$ strip of the triangular
lattice with cyclic or M\"obius boundary conditions. For comparison, zeros of
the flow polynomial $F(tri,3 \times L_x,cyc.,q)$ for $L_x=20$ (so that this
polynomial has degree 81) are also shown.}}
\label{tpxy3flow}
\end{figure}

The locus ${\cal B}$ for the cyclic and M\"obius $L_y=3$ strips of the
triangular lattice is shown in the $q$ plane in Fig. \ref{tpxy3flow}.  
This locus is noncompact in the $q$
plane, containing eight curves that extend infinitely far away from $q=0$.
The locus separates the $q$ plane into several regions. Three of these
regions, $R_j$, $j=1,2,3$, contain interval of the real axis, which are $q
\ge q_{cf}$ for $R_1$, and $2 \le q \le q_{cf}$ for $R_2$, and $q < 2$ for
$R_3$, where 
\beq
q_{cf} = 2.213548 \quad {\rm for} \quad tri, 3 \times \infty,cyc./Mb. \ . 
\label{qctripxy3}
\eeq
This is within about 15 \% of the asymptotic value $q_{cf} \simeq 2.618$ for
the 2D triangular lattice (see eq. (\ref{qcftri}) below). In regions $R_1$ and
$R_3$, the dominant term is the root of maximal magnitude of the cubic equation
(\ref{trieq33}), which we denote $\lambda_{tri,3,1,jmax}$, so that
\beq
fl(tri[3 \times \infty,cyc./Mb.],q)=(\lambda_{tri,3,1,jmax})^{1/4} \quad 
{\rm for} \quad q \in R_1 . 
\label{ftripxy3region1}
\eeq
In region $R_2$, $\lambda_{tri,3,2,2}$ is dominant, so that 
\beq
|fl(tri[3 \times \infty,cyc./Mb.],q)|=|\lambda_{tri,3,2,2}|^{1/4} \quad 
{\rm for} \quad q \in R_2 \ .
\label{ftripxy3region2}
\eeq
In region $R_3$, 
\beq
|fl(tri[3 \times \infty,cyc./Mb.],q)|=|\lambda_{tri,3,1,jmax}|^{1/4} \quad
{\rm for} \quad q \in R_3 . 
\label{ftripxy3region3}
\eeq
In addition to the regions $R_j$, $j=1,2,3$ that contain intervals of the real
axis, there are also three complex-conjugate pairs of regions away from the
real axis, $R_j$, $R_j^*$, $j=4,5,6$.  These can be identified in
Fig. \ref{tpxy3flow} as follows: $R_4$ contains the point $q=4 + 3i$ and
extends to complex infinity; $R_5$ contains the point $q=1.5+3i$ and extends to
complex infinity; and $R_6$ contains the point $q=-1+3i$ and extends to complex
infinity.

\section{$L_{\lowercase{y}}=2$ Strip of the Triangular Lattice with Toroidal 
and Klein Bottle Boundary Conditions}

For the $L_y=2$ strips of the triangular lattice with torus and Klein bottle
boundary conditions, we find that 
\beqs
& & \lambda_{tri2t,j}=\frac{1}{2}\biggl [ 11-19q+15q^2-6q^3+q^4 \cr\cr
& & \pm (129-446q+727q^2-722q^3+479q^4-218q^5+66q^6-12q^7+q^8)^{1/2} \
\biggr ] \quad j=1,2 \cr\cr
& & 
\label{flamtpxpy2_12}
\eeqs
\beqs
& & \lambda_{tri2t,j}=\frac{1}{2}\biggl [ 14-20q+15q^2-6q^3+q^4 \cr\cr
& & \pm (212-600q+852q^2-776q^3+493q^4-220q^5+66q^6-12q^7+q^8)^{1/2} \ 
\biggr ] \quad j=3,4 \cr\cr
& & 
\label{flamtpxpy2_34}
\eeqs
and 
\beq
\lambda_{tri2t,5}=2 \ .
\label{flamtpxpy2_5}
\eeq
The corresponding coefficients are
\beq
c_{tri2t,j}=1 \quad j=1,2
\label{ct2t_12}
\eeq
\beq
c_{tri2t,j}=q-1 \quad j=3,4
\label{ct2t_34}   
\eeq
and
\beq
c_{tri2t,5}=\frac{1}{2}q(q-3) \ .
\label{ct2t_5}   
\eeq
It follows that 
\beqs
& & F(tri[2 \times m,torus],q)=(\lambda_{tri2t,1})^m +
(\lambda_{tri2t,2})^m \cr\cr
& & + (q-1)\Bigl [ (\lambda_{tri2t,3})^m + (\lambda_{tri2t,4})^m \Bigr ] + 
\frac{1}{2}q(q-3) 2^m \ .
\label{flowtripxpy2}
\eeqs

\begin{figure}[hbtp]
\centering
\leavevmode
\epsfxsize=4.0in
\begin{center}
\leavevmode
\epsffile{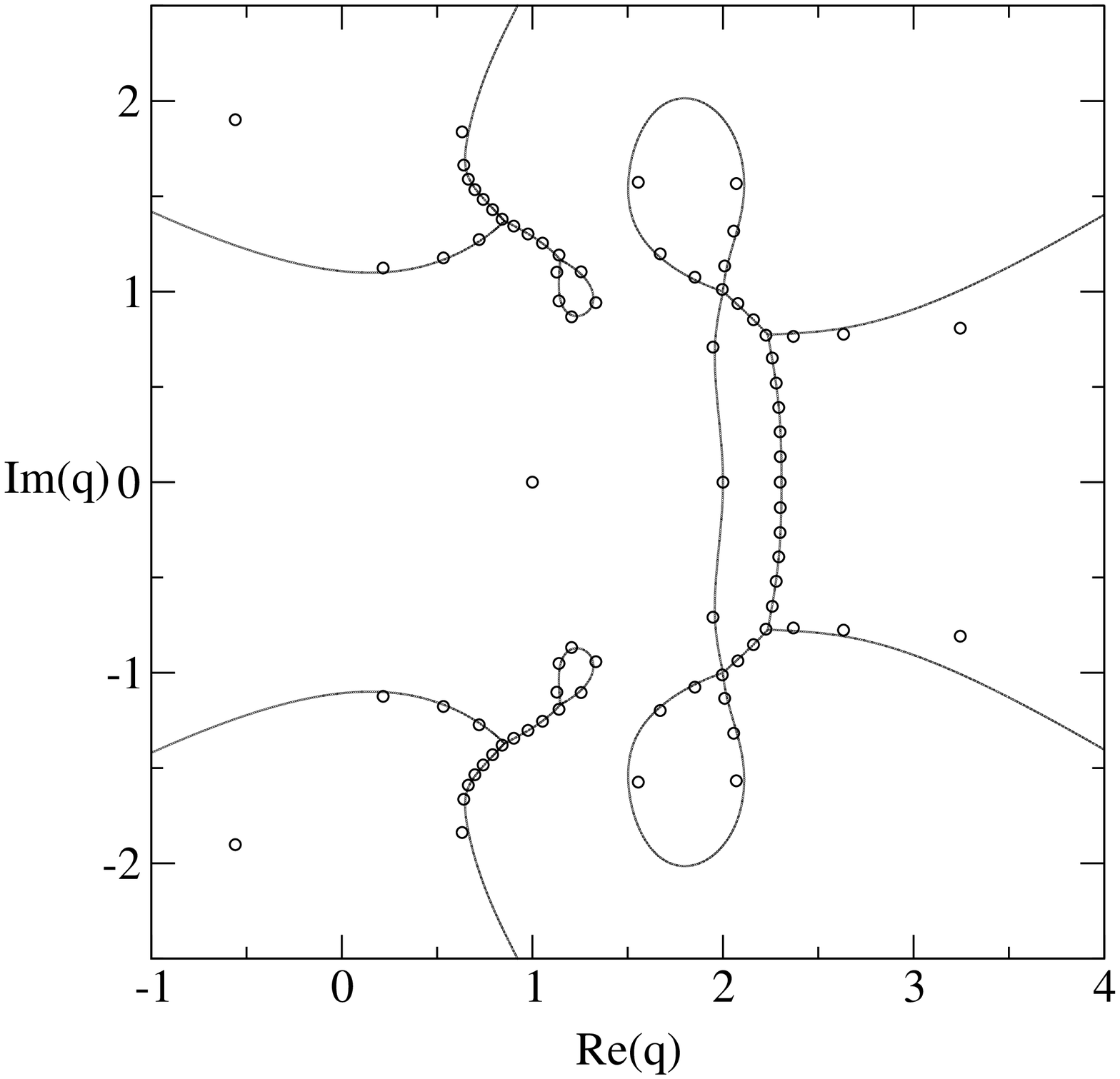}
\end{center}
\caption{\footnotesize{Singular locus ${\cal B}$ in the $q$ plane for $fl(tri,2
\times \infty,torus/Kb.,q)$ for the $2 \times \infty$ strip of the triangular
lattice with torus or Klein bottle boundary conditions. For comparison, zeros
of the flow polynomial $F(tri,2 \times L_x,torus.,q)$ for $L_x=20$ are also
shown.}}
\label{tpxpy2flow}
\end{figure}
 
\begin{figure}[hbtp]
\centering
\leavevmode
\epsfxsize=4.0in
\begin{center}
\leavevmode
\epsffile{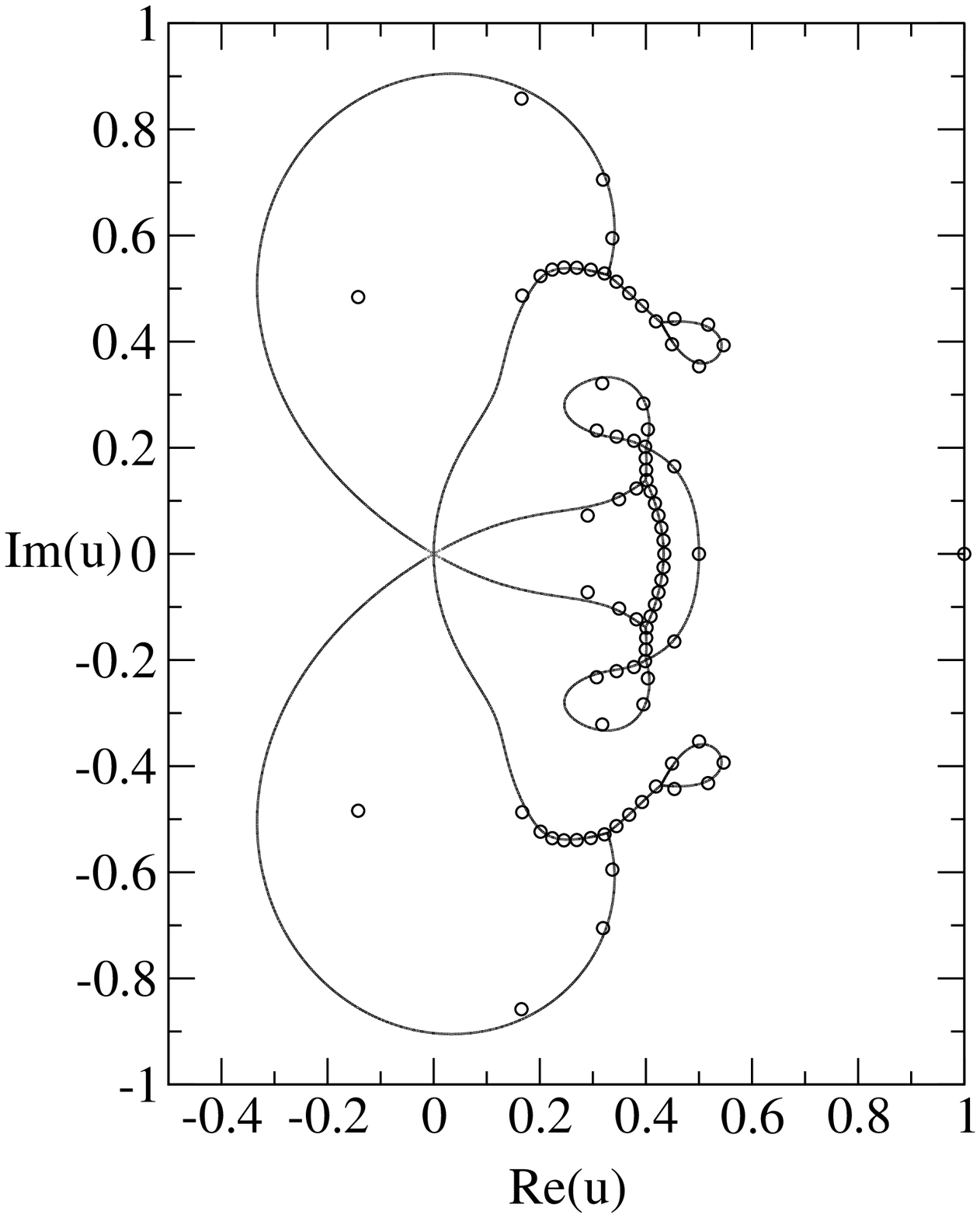}
\end{center}
\caption{\footnotesize{Singular locus ${\cal B}$ in the $u$ plane for $fl(tri,2
\times \infty,torus/Kb.,q)$ for the $2 \times \infty$ strip of the triangular
lattice with torus or Klein bottle boundary conditions. For comparison, zeros
of the flow polynomial $F(tri,2 \times L_x,torus,q)$ for $L_x=20$, expressed
in terms of $u$, are also shown.}}
\label{tpxpy2uflow}
\end{figure}

The locus ${\cal B}$ for the torus and Klein bottle $L_y=2$ strips of the
triangular lattice is shown in the $q$ plane in Fig. \ref{tpxpy2flow} and
in the
$u$ plane in Fig. \ref{tpxpy2uflow}.  This locus is noncompact in the $q$
plane, and divides this plane into various regions. Three of these
regions, $R_j$, $j=1,2,3$, contain intervals of the real axis, which are
$q \ge q_{cf}$ for $R_1$, $2 \le q \le q_{cf}$ for $R_2$, and $q < 2$ for
$R_3$, where
\beq
q_{cf} = 2.307144568... \quad {\rm for} \quad tri, 2 \times \infty,torus/Kb.
\ . 
\label{qctripxpy2}
\eeq
The dominant terms in regions $R_j$, $j=1,2,3$ are $\lambda_{tri2t,1}$,
$\lambda_{tri2t,5}$ and $\lambda_{tri2t,3}$ so that

\beq
fl(tri[2 \times \infty, torus/Kb.],q)=
(\lambda_{tri2t,1})^{1/4} \quad {\rm for} \quad q \in R_1
\label{ftripxpy2region1}
\eeq

\beq
|fl(tri[2 \times \infty, torus/Kb.],q)|=
2^{1/4} \quad {\rm for} \quad q \in R_2
\label{ftripxpy2region2}
\eeq

\beq
|fl(tri[2 \times \infty, torus/Kb.],q)|=                               
|\lambda_{tri2t,3}|^{1/4} \quad {\rm for} \quad q \in R_3 \ .
\label{ftripxpy2region3}
\eeq
There are also three complex-conjugate pairs of regions, $R_j$, $R_j^*$,
$j=4,5,6$. These can be identified in Fig. \ref{tpxpy2flow} as follows:
$R_4$ is a ``bubble'' region centered at the point $q=1.8 + 1.5i$; $R_5$
is a small ``bubble'' region centered at the point $q=1.2 + i$; and $R_6$
contains the point $q=2i$ and extends to complex infinity.  The 
dominant terms in regions $R_j$, $j=4,5,6$ are $\lambda_{tri2t,1}$,
$\lambda_{tri2t,5}$ and $\lambda_{tri2t,1}$.

To show that ${\cal B}$ passes through $u=0$, we calculate the corresponding
scaled terms $\lambda_{tri2t,j,u} = \lambda_{tri2t,j}/q^4$ with $u=1/q$ and see
that the degeneracy condition of dominant $\lambda_{tri2t,j,u}$'s has a
solution at $u=0$.  The dominant terms in the vicinity of $u=0$ are
$\lambda_{tri2t,1,u}$ and $\lambda_{tri2t,3,u}$.
For $u \to 0$, these terms have the Taylor series expansion 
\beq
\lambda_{tri2t,1,u}=1-6u+15u^2-19u^3+O(u^4) 
\label{flamtri21utaylor}
\eeq
\beq
\lambda_{tri2t,3,u}=1-6u+15u^2-20u^3+O(u^4) \ .
\label{flamtri23utaylor}
\eeq
Hence, introducing polar coordinates $u=re^{i\theta}$, the condition of
degeneracy of magnitudes in the vicinity of $u=0$ yields the condition
\beq
r^3 \cos \theta (3-4\cos^2\theta) = 0 \quad {\rm as} \quad r \to 0 \ .
\label{polareq2}
\eeq
This proves that four branches of ${\cal B}$ approach the origin of the $u$ 
plane at the angles
\beq
\theta = \frac{(2k+1)\pi}{6} \ , \quad {\rm for} \quad 1 \le k \le 6
\label{thetasoltripxpy2}
\eeq
i.e., at $\theta=\pm \pi/6$, $\theta=\pm \pi/2$ and $\theta=\pm 5\pi/6$.

\section{Strips of the Triangular Lattice with Free Boundary Conditions}

In this section, we give the flow polynomials of the triangular lattice 
with free boundary conditions.  First, we have 
\beq
F(tri[L_x=2,L_x=m,free],q)=(q-1)(q-2)^{2(m-1)+1} \ .
\label{flowtrixy2}
\eeq
For the cases $L_y \ge 3$, we give the result in term of generating 
functions as shown from eq. (\ref{gammazcyl}) to eq. (\ref{d}). For 
$F(tri[L_y=3,L_x=m,free],q)$ we calculate 
\beq
{\cal N}(tri[L_y=3,free],q,z) = (q-1)(q-2)^2z[(q-2) + 
(q-1)^3z - (q-1)^2z^2]
\label{trifreely3n}
\eeq
\beq
{\cal D}(tri[L_y=3,free],q,z) = 1 - (q^4-7q^3+21q^2-33q+23)z 
+ 2(q-2)^2z^2 \ .
\label{trifreely3d}
\eeq
There are thus $N_{F,tri,L_y=3,free,\lambda}=2$ terms, 
\beq
\lambda_{tri,L_y=3,free,j}=\frac{1}{2}\biggl [ q^4-7q^3+21q^2-33q+23 \pm 
\sqrt{R_{t3f}} \ \biggr ] \ , \quad j=1,2 
\label{lamty3j}
\eeq
where
\beq
R_{t3f}=q^8-14q^7+91q^6-360q^5+949q^4-1708q^3+2047q^2-1486q+497 \ . 
\label{rt3f}
\eeq

\begin{figure}[hbtp]
\centering
\leavevmode
\epsfxsize=4.0in
\begin{center}
\leavevmode
\epsffile{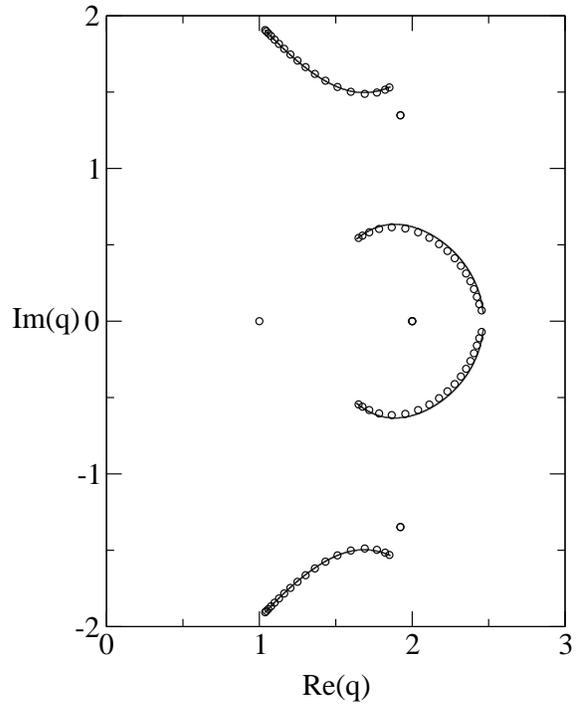}
\end{center}
\caption{\footnotesize{Singular locus ${\cal B}$ in the $q$ plane for $fl(tri,3
\times \infty,free,q)$ for the $3 \times \infty$ strip of the triangular
lattice with free boundary conditions. For comparison, zeros
of the flow polynomial $F(tri,3 \times L_x,free,q)$ for $L_x=21$ are also
shown.}}
\label{ty3flow}
\end{figure}

The locus ${\cal B}$ for the $L_x \to \infty$ limit of this strip is shown in
Fig. \ref{ty3flow}.  This locus is compact, consisting of two pairs of
complex-conjugate arcs that do not separate the $q$ plane into different
regions and do not cross the real axis, so that no $q_{cf}$ is defined.  
The eight endpoints of the arcs occur at the eight zeros of $R_{t3f}$, where
the square root in (\ref{lamty3j}) have branch point singularities.  Thus, we
find that the loci ${\cal B}$ for all of the lattice strips that we have
considered with free longitudinal boundary conditions (which include the strips
with free and cylindrical boundary conditions) are compact.  These thus
contrast with the behavior of ${\cal B}$ for many of the strips with periodic
or twisted periodic longitudinal boundary conditions, for which ${\cal B}$ is
noncompact. 

For $F(tri[L_y=4,L_x=m,free],q)$ we find $N_{F,tri,L_y=4,free,\lambda}=6$
and
the following coefficients of the generating function:
\beq
A_{tri,4,free,0} = 0
\label{sqfreely4a1}
\eeq
\beq
A_{tri,4,free,1} = (q-1)(q-2)^5
\label{trifreely4a1}
\eeq
\beq
A_{tri,4,free,2} = (q-1)(q-2)^3(2q^6-18q^5+69q^4-140q^3+152q^2-80q+17)
\label{trifreely4a2} 
\eeq
\beqs
A_{tri,4,free,3} & = & 
(q-1)(q-2)^3(q^{10}-14q^9+89q^8-334q^7+811q^6-1328q^5 \cr\cr
& & +1520q^4-1314q^3+962q^2-557q+160)
\label{trifreely4a} 
\eeqs
\beqs
A_{tri,4,free,4} & = & 
-(q-1)^2(q-2)^3(q^{11}-15q^{10}+102q^9-404q^8+994q^7-1452q^6 \cr\cr
& & +856q^5+990q^4-2580q^3+2406q^2-1100q+194)
\label{trifreely4a4} 
\eeqs
\beqs
& & A_{tri,4,free,5} =  
(q-1)^3(q-2)^3(2q^{11}-37q^{10}+319q^9-1678q^8+5936q^7 \cr\cr
& & -14675q^6+25474q^5-30225q^4+22685q^3-8607q^2+9q+793)
\label{trifreely4a5} 
\eeqs
\beqs
A_{tri,4,free,6} & = & 
(q-1)^4(q-2)^3(q^{12}-24q^{11}+265q^{10}-1785q^9+8181q^8-26897q^7 \cr\cr
& & +65062q^6-116663q^5+153869q^4-145558q^3+93785q^2-37025q \cr\cr
& & +6809)
\label{trifreely4a6} 
\eeqs
\beq
A_{tri,4,free,7} = (q-1)^6(q-2)^3(2q^5-19q^4+75q^3-151q^2+155q-65)
\label{trifreely4a7} 
\eeq
\beq
b_{tri,4,free,1} = -q^6+10q^5-46q^4+126q^3-221q^2+240q-129   
\label{trifreely4b1}
\eeq
\beqs
b_{tri,4,free,2} & = & 
q^{10}-18q^9+150q^8-760q^7+2592q^6-6230q^5+10746q^4 \cr\cr
& & -13278q^3+11457q^2-6404q+1820
\label{trifreely4b2} 
\eeqs
\beqs
b_{tri,4,free,3} & = & 
-q^{12}+22q^{11}-227q^{10}+1448q^9-6347q^8+20118q^7-47275q^6 \cr\cr
& & +83064q^5-108597q^4+103580q^3-69056q^2+29332q-6141
\label{trifreely4b3} 
\eeqs
\beqs
& & b_{tri,4,free,4} =  
2q^{12}-44q^{11}+450q^{10}-2824q^9+12099q^8-37264q^7+84596q^6 \cr\cr
& & -142676q^5+177551q^4-159174q^3+97739q^2-36996q+6557
\label{trifreely4b4} 
\eeqs
\beqs
b_{tri,4,free,5} & = & 
-(q-1)^2(q-2)^2(q^8-16q^7+114q^6-470q^5+1225q^4-2066q^3 \cr\cr
& & +2201q^2-1356q+375)
\label{trifreely4b5} 
\eeqs
\beq
b_{tri,4,free,6} = (q-1)^4(q-2)^4 \ .
\label{trifreely4b6} 
\eeq

We have also calculated the generating function for 
$F(tri[L_y=5,L_x=m,free],q)$.  The results are too lengthy to present
here, but
we mention that 
\beq
{\rm deg} \ {\cal D}(tri,L_y=5,free) = N_{F,tri,5,free} = 13 \ .
\label{ntottrifree5}
\eeq
The respective loci ${\cal B}$ for the $L_x \to \infty$ limits of these strips
of the triangular lattice with $L_y=4,5$ and free boundary conditions again
consist of arcs.

\section{Strips of the Triangular Lattice with Cylindrical Boundary 
Conditions}

We give the result in term of generating functions as shown from eq. 
(\ref{gammazcyl}) to eq. (\ref{d}). For 
$F(tri[L_y=2,L_x=m,cyl.],q)$ we have 
\beq
{\cal N}(tri[L_y=2,cyl.],q,z) = (q-1)[1 - (q-1)(q-2)(q-3)z]
\label{trcylly2n}
\eeq
\beq
{\cal D}(tri[L_y=2,cyl.],q,z) = 1 - (q^4-6q^3+15q^2-19q+11)z 
- (q-1)^3(q-2)z^2 \ .
\label{tricylly2d}
\eeq

The coefficients of the generating function for
$F(tri,L_y=3,L_x=m,cyl.,q)$ are
\beq
A_{tri,3,cyl,0} = q-1
\label{tricylly3a0}
\eeq
\beq
A_{tri,3,cyl,1} = -(q-1)^2(2q^4-17q^3+58q^2-97q+68)
\label{tricylly3a1}
\eeq
\beq
A_{tri,3,cyl,2} = -(q-1)^4(q-2)(q^3-3q^2-4q+13)
\label{tricylly3a2}
\eeq
\beq
b_{tri,3,cyl,1} = -q^6+9q^5-36q^4+84q^3-127q^2+125q-65
\label{tricylly3b1}
\eeq
\beq
b_{tri,3,cyl,2} = (q-1)(q^7-11q^6+49q^5-110q^4+119q^3-24q^2-69q+52)
\label{tricylly3b2}
\eeq
\beq
b_{tri,3,cyl,3} = -(q-1)^5(q-2)(q^2-2q-1) \ .
\label{tricylly3b3}
\eeq

We have also calculated the generating function for 
$F(tri[L_y=4,L_x=m,cyl.],q)$.  The results are too lengthy to present
here, but
we mention that 
\beq
{\rm deg} \ {\cal D}(tri,L_y=4,cyl.) = N_{F,tri,4,cyl} = 6 \ .
\label{ntottricyl4}
\eeq

\section{${\cal B}$ for 2D Lattices}

We can also obtain some results on ${\cal B}_{fl}$, in particular, $q_{cf}$
values, for (infinite) 2D lattices.  These follow directly from
eqs. (\ref{bflwdual}).  Using the relation (\ref{bflwsq})
together with the result that $q_c(sq)=3$ \cite{lenard}, we have
\beq
q_{cf}(sq) =  3 \ .
\label{qcfsq}
\eeq
Using (\ref{bflwtrihc}) together with the result that $q_c(tri)=4$
\cite{baxter}, we have
\beq
q_{cf}(hc)=4 \ .
\label{qcfhc}
\eeq
For the honeycomb lattice, formally, $q_c=(1/2)(3+\sqrt{5})$ \cite{ss,p3afhc},
so that
\beq
q_{cf}(tri) = \frac{3+\sqrt{5}}{2} = 2.6180...
\label{qcftri}
\eeq
Since $q_c(kag)=3$, we finally have
\beq
q_{cf}(diced)=3 \ .
\label{qcdiced}
\eeq

As was discussed in our previous works on chromatic polynomials, from a study
of the loci ${\cal B}_W$ for infinite-length strips of regular lattices of
increasing widths (and with a variety of boundary conditions), one can
formulate plausible inferences about the the boundary ${\cal B}_W$ for the
limit of infinite width, i.e., the full 2D lattice.  Given the duality relation
(\ref{bflwdual}), one can use either ${\cal B}_W$ or ${\cal B}_{fl}$, or both,
for this purpose.  This program is useful since one does not know ${\cal B}_W$
for any 2D lattice except for results on the triangular lattice (\cite{baxter},
see, however, \cite{sstri}). It was found in \cite{w} and subsequent papers
that if one uses periodic longitudinal boundary conditions, then ${\cal B}_W$
exhibits a number of features that one would infer to hold for the 2D lattice,
such as the property of separating the $q$ plane into various regions, and
crossing the real $q$ axis at $q=0$ and a maximal value, $q_c$.  The
compactness of the loci ${\cal B}_W$ found for these strips is also the same as
the compactness property of ${\cal B}_W$ for infinite regular lattices.  This
latter property is established by taking the $|V| \to \infty$ limit of a
section of regular lattice $\Lambda$ and applying the bound (\ref{sokal}). This
compactness is in accord with the fact, as discussed in \cite{wa,wa2}, that if
${\cal B}_W$ did extend infinitely far away from the origin, passing through
$1/q=0$, then this point would be a point of nonanalyticity of the function
$W(\Lambda,q)/q$, thereby precluding a Taylor-series expansion of this function
around this point.  However, well-known procedures exist for calculating
Taylor-series expansions of $W(\Lambda,q)/q$ around $1/q=0$ for regular
lattices \cite{nagle,baker}.

In contrast, as discussed above, we find that the loci ${\cal B}_{fl}$ for the
infinite-length limits of strips with periodic longitudinal boundary conditions
are usually noncompact (an exception being the case of self-dual strips of the
square lattice) and do not, in general, pass through $q=0$, although they do
separate the $q$ plane into various regions.
 From the duality relation (\ref{bflwdual}) in conjunction with the bound
(\ref{sokal}), we infer that ${\cal B}_{fl}$ is compact for any infinite
regular planar lattice $\Lambda$ that has a dual planar lattice $\Lambda^*$
(given the fact that a regular lattice has a fixed, finite degree for all of
its vertices).  In view of this result, there are thus two possibilities: (i)
either ${\cal B}_{fl}$ will become compact for sufficiently great finite width
for each type of infinite-length lattice strip, or (ii) we encounter another
kind of noncommutativity in addition to (\ref{flnoncom}), namely that
$\lim_{L_y \to \infty} {\cal B}_{fl}(G_s,L_y \times \infty)$ is different from
the accumulation set of the zeros of the flow polynomial of an $L_y \times L_x$
section of a regular lattice obtained by letting $L_x$ and $L_y$ both approach
infinity with $L_y/L_x$ a finite nonzero constant.  Further study is needed to
decide which of these two types of behavior occurs.

Our present results show that the use of ${\cal B}_W$ and ${\cal B}_{fl}$
calculated on infinite-length finite-width strips are somewhat complementary.
For example, the study of the loci ${\cal B}_W$ for finite-width,
infinite-length strips of a lattice $\Lambda$ with periodic longitudinal
boundary conditions has the advantage that these loci are compact, as is true
of ${\cal B}_W(\Lambda)= {\cal B}_{fl}(\Lambda^*)$ for the infinite dual pair
of lattices $\Lambda$ and $\Lambda^*$.  On the other hand, the loci ${\cal
B}_{fl}$ for all of the cyclic/M\"obius strips of the square lattice for which
we have calculated them, do exhibit the interesting feature of having $q_{cf}$
equal to the value 3 for the infinite square lattice, whereas the $q_c$ values
for ${\cal B}_W$ on these strips only approach the square-lattice value
asymptotically as $L_y$ increases.  Of course, for the self-dual families of
planar strip graphs studied in \cite{dg,sdg}, ${\cal B}_{fl}={\cal B}_W$ for
each value of $L_y$ separately, and in these cases, these loci share the common
property of being compact and having $q_c=q_{cf}=3$, the asymptotic value.

\section{Summary} 

In this paper we have given exact calculations of flow polynomials $F(G,q)$ for
lattice strips of various fixed widths and arbitrarily great lengths, with
several different boundary conditions.  We have determined the resultant
functions $fl$ giving the $q$-flows per face in the limit of infinite-length
strips.  We have also studied the zeros of $F(G,q)$ in the complex $q$ plane
and determined exactly the asymptotic accumulation set of these zeros ${\cal
B}_{fl}$, across which $fl$ is nonanalytic in the infinite-length limit.  We
found that these loci were noncompact for many strip graphs with periodic (or
twisted periodic) longitudinal boundary conditions, in contrast to the
usual behavior for the analogous loci ${\cal B}_W$ for the $W$ function
obtained from chromatic polynomials for these strips.  We also found the
interesting feature that, aside from the trivial case $L_y=1$, the maximal
point, $q_{cf}$, where ${\cal B}$ crosses the real axis, is universal on cyclic
and M\"obius strips of the square lattice for all widths for which we have
calculated it and is equal to the asymptotic value $q_{cf}=3$ for the infinite
square lattice.  Duality relations were used to derive a number of connections
between $fl$ and $W$, and ${\cal B}_{fl}$ and ${\cal B}_W$, for planar families
of graphs.

\bigskip
\bigskip

Acknowledgments: 

It is a pleasure to dedicate this paper to Professor F. Y. Wu, who has
contributed so much to statistical mechanics and, in particular, to the
understanding of the Potts model, on the occasion of his 70'th birthday.
This research was partially supported by the NSF grant PHY-97-22101.

\bigskip
\bigskip

\section{Appendix}

\subsection{$L_{\lowercase{y}}=2$ Cyclic and M\"obius Strips of the Square 
Lattice}

In this appendix we give some explicit flow polynomials for the families of
graphs considered in the text, in order to illustrate the general discussion in
the text. We begin with strips of the square lattice.  For $L_y=2$ we have
\beq
F(sq[2 \times 1,cyc.],q) = 0
\label{flowsq2x1cyc}
\eeq
(this vanishes since this graph contains a bridge)
\beq
F(sq[2 \times 2,cyc.],q) = (q-1)(q-2)^2
\label{flowsq2x2cyc}
\eeq
\beq
F(sq[2 \times 3,cyc.],q) = (q-1)(q-2)(q-3)^2
\label{flowsq2x3cyc}
\eeq
\beq
F(sq[2 \times 4,cyc.],q) = (q-1)(q-2)(q^3-9q^2+29q-32)
\label{flowsq2x4cyc}
\eeq
\beq
F(sq[2 \times 5,cyc.],q) = (q-1)(q-2)(q-3)(q^3-9q^2+30q-35)
\label{flowsq2x5cyc}
\eeq
\beq
F(sq[2 \times 1,Mb.],q) = (q-1)(q-2)
\label{flowsq2x1mb}
\eeq
\beq
F(sq[2 \times 2,Mb.],q) = (q-1)(q-2)(q-3)
\label{flowsq2x2mb}
\eeq
\beq
F(sq[2 \times 3,Mb.],q) = (q-1)(q-2)(q^2-6q+10)
\label{flowsq2x3mb}
\eeq
\beq
F(sq[2 \times 4,Mb.],q) = (q-1)(q-2)(q-3)(q^2-6q+11)
\label{flowsq2x4mb}
\eeq
\beq
F(sq[2 \times 5,Mb.],q) = (q-1)(q-2)(q^4-12q^3+57q^2-125q+106)
\label{flowsq2x5mb}
\eeq

\subsection{$L_{\lowercase{y}}=3$ Cyclic and M\"obius Strips of the Square
Lattice} 

\beq F(sq[3 \times 1,cyc.],q) = 0
\label{flowsq3x1cyc}
\eeq
(where again, this vanishes because the graph contains a bridge)
\beq
F(sq[3 \times 2,cyc.],q) = (q-1)(q-2)^2(q^2-3q+3)
\label{flowsq3x2cyc}
\eeq
\beq
F(sq[3 \times 3,cyc.],q) = (q-1)(q-2)(q-3)^2(q^3-6q^2+14q-13)
\label{flowsq3x3cyc}
\eeq
\beqs
F(sq[3 \times 4,cyc.],q) & = & (q-1)(q-2)(q^7-17q^6+129q^5-569q^4+1588q^3
\cr\cr
& & -2824q^2+2969q-1416)
\label{flowsq3x4cyc}
\eeqs
\beqs
F(sq[3 \times 5,cyc.],q) & = &
(q-1)(q-2)(q-3)(q^8-19q^7+165q^6-860q^5+2964q^4
\cr\cr
& & -6977q^3+11034q^2-10740q+4895)
\label{flowsq3x5cyc}
\eeqs
\beq
F(sq[3 \times 1,Mb.],q) = (q-1)^2(q-2)
\label{flowsq3x1mb}
\eeq
\beq
F(sq[3 \times 2,Mb.],q) = (q-1)(q-2)(q^3-7q^2+18q-17)
\label{flowsq3x2mb}
\eeq
\beq
F(sq[3 \times 3,Mb.],q) = (q-1)(q-2)(q^5-12q^4+61q^3-167q^2+253q-172) 
\label{flowsq3x3mb}
\eeq
\beqs
F(sq[3 \times 4,Mb.],q) & = & (q-1)(q-2)(q^7-17q^6+129q^5-571q^4 \cr\cr
& & +1610q^3-2930q^2+3222q-1655)
\label{flowsq3x4mb}
\eeqs

\subsection{$L_{\lowercase{y}}=4$ Cyclic and M\"obius Strips of the Square
Lattice} 

\beq F(sq[4 \times 1,cyc.],q) = 0
\label{flowsq4x1cyc}
\eeq
\beq
F(sq[4 \times 2,cyc.],q) = (q-1)(q-2)^2(q^2-3q+3)^2
\label{flowsq4x2cyc}
\eeq
\beq
F(sq[4 \times 3,cyc.],q) = (q-1)(q-2)(q-3)^2(q^3-6q^2+14q-13)^2
\label{flowsq4x3cyc}
\eeq
\beqs
& & F(sq[4 \times 4,cyc.],q) = (q-1)(q-2)
(q^{11}-25q^{10}+293q^9-2128q^8+10664q^7-38808q^6 \cr\cr
& & +104899q^5-211073q^4+310424q^3-318191q^2+204597q-62416)
\label{flowsq4x4cyc}
\eeqs
\beq
F(sq[4 \times 1,Mb.],q) = (q-1)(q-2)(q^2-3q+3)
\label{flowsq4x1mb}
\eeq
\beq
F(sq[4 \times 2,Mb.],q) = (q-1)(q-2)(q^5-11q^4+52q^3-133q^2+188q-119)
\label{flowsq4x2mb}
\eeq
\beqs
F(sq[4 \times 3,Mb.],q) & = & 
(q-1)(q-2)(q^8-18q^7+148q^6-730q^5+2380q^4-5301q^3 \cr\cr
& & +7968q^2-7475q+3369)
\label{flowsq4x3mb}
\eeqs
\beqs
& & F(sq[4 \times 4,Mb.],q) = 
(q-1)(q-2)(q^{11}-25q^{10}+293q^9-2131q^8+10724q^7-39374q^6 \cr\cr
& & +108174q^5-223727q^4+343714q^3-376152q^2+265280q-91429)
\label{flowsq4x4mb}
\eeqs

\subsection{$L_{\lowercase{y}}=2$ Strips of the Square Lattice with Torus and 
Klein Bottle Boundary Conditions}

\beq
F(sq[2 \times 1,torus],q) = (q-1)^3
\label{flowsqtorus2x1}
\eeq
\beq
F(sq[2 \times 2,torus],q) = (q-1)(q^4-7q^3+21q^2-29q+15)
\label{flowsqtorus2x2}
\eeq
\beq
F(sq[2 \times 3,torus],q) = (q-1)(q^6-11q^5+55q^4-156q^3+262q^2-244q+97)
\label{flowsqtorus2x3}
\eeq
\beqs
F(sq[2 \times 4,torus],q) & = & (q-1)(q^8-15q^7+105q^6-441q^5+1205q^4
\cr\cr
& & -2182q^3+2550q^2-1755q+543)
\label{flowsqtorus2x4}
\eeqs
\beq
F(sq[2 \times 1,kb],q) = (q-1)(q^2-3q+3)
\label{flowsqklein2x1}
\eeq
\beq
F(sq[2 \times 2,kb],q) = (q-1)(q^4-7q^3+21q^2-30q+17)
\label{flowsqklein2x2}
\eeq
\beq
F(sq[2 \times 3,kb],q) = (q-1)(q^6-11q^5+55q^4-156q^3+262q^2-245q+99)
\label{flowsqklein2x3}
\eeq
\beqs
F(sq[2 \times 4,kb],q) & = & (q-1)(q^2-4q+5)(q^6-11q^5+56q^4-162q^3 \cr\cr
& & +277q^2-264q+109)
\label{flowsqklein2x4}
\eeqs

\subsection{$L_{\lowercase{y}}=3$ Strips of the Square Lattice with Torus
and Klein Bottle Boundary Conditions}

\beq 
F(sq[3 \times 1,torus],q) = (q-1)^4
\label{flowsq3torus2x1}
\eeq
\beq
F(sq[3 \times 2,torus],q) = (q-1)(q^6-11q^5+55q^4-156q^3+262q^2-244q+97)
\label{flowsq3torus2x2}
\eeq
\beqs
F(sq[3 \times 3,torus],q) & = &
(q-1)(q^9-17q^8+136q^7-671q^6+2254q^5-5363q^4+9076q^3 \cr\cr
& & -10564q^2+7655q-2605)
\label{flowsq3torus2x3}
\eeqs
\beqs
F(sq[3 \times 4,torus],q) & = &
(q-1)(q^{12}-23q^{11}+253q^{10}-1759q^9+8615q^8-31363q^7 \cr\cr
& & +87233q^6-187346q^5+309415q^4-384602q^3+342441q^2 \cr\cr
& & -196170q+54481)
\label{flowsq3torus2x4}
\eeqs
\beq
F(sq[3 \times 1,kb],q) = (q-1)^2(q^2-3q+3)
\label{flowsq3klein2x1}  
\eeq
\beq
F(sq[3 \times 2,kb],q) = (q-1)(q^6-11q^5+55q^4-158q^3+275q^2-273q+119)
\label{flowsq3klein2x2}
\eeq
\beqs
F(sq[3 \times 3,kb],q) & = &
(q-1)(q^9-17q^8+136q^7-671q^6+2254q^5-5365q^4+9096q^3 \cr\cr
& & -10645q^2+7808q-2715)
\label{flowsq3klein2x3}
\eeqs
\beqs
F(sq[3 \times 4,kb],q) & = &
(q-1)(q^{12}-23q^{11}+253q^{10}-1759q^9+8615q^8-31363q^7 \cr\cr
& & +87233q^6-187348q^5+309441q^4-384752q^3+342906q^2 \cr\cr
& & -196915q+54959)
\label{flowsq3klein2x4}
\eeqs

\subsection{$L_{\lowercase{y}}=4$ Strips of the Square Lattice with 
Cylindrical Boundary Conditions}

\beq
F(sq[4 \times 1,cyl],q)=q-1
\label{flowsqcyl4x1}
\eeq
\beq
F(sq[4 \times 2,cyl],q)=F(sq,2 \times 4,cyc,q)=(q-1)(q-2)(q^3-9q^2+29q-32)
\label{flowsqcyl4x2}
\eeq
\beqs
& & F(sq[4 \times 3,cyl],q)=(q-1)(q-2)(q^7-17q^6+129q^5-569q^4 \cr\cr
& & +1588q^3-2824q^2+2969q-1416)
\label{flowsqcyl4x3}
\eeqs
\beqs
& & F(sq[4 \times 4,cyl],q)=(q-1)(q-2)(q^{11}-25q^{10}+293q^9-2128q^8
\cr\cr
& & +10664q^7-38808q^6+104899q^5-211073q^4+310424q^3-318191q^2 \cr\cr
& & +204597q-62416)
\label{flowsqcyl4x5}
\eeqs

\subsection{$L_{\lowercase{y}}=3$ Cyclic and M\"obius Strips of the Honeycomb 
Lattice}

\beq
F(hc[3 \times 2,cyc.],q)=(q-1)(q-3)(q-2)^3
\label{flowhc3x2cyc}
\eeq
\beq
F(hc[3 \times 3,cyc.],q)=(q-1)(q-2)(q-3)^2(q^3-9q^2+29q-32)
\label{flowhc3x3cyc}
\eeq
\beqs
F(hc[3 \times 4,cyc.],q) & = & (q-1)(q-2)(q-3)(q^6-18q^5+141q^4-620q^3
\cr\cr
& & +1619q^2-2368q+1496)
\label{flowhc3x4cyc}
\eeqs
\beqs
& & F(hc[3 \times
5,cyc.],q)=(q-1)(q-2)(q-3)(q^8-24q^7+260q^6-1670q^5+6999q^4
\cr\cr & & -19698q^3+36408q^2-40240q+20170)
\label{flowhc3x5cyc}
\eeqs

\subsection{$L_{\lowercase{y}}=2$ Cyclic Strips of the 
Triangular Lattice}

\beq
F(tri[2 \times 1,cyc.],q) = (q-1)^3
\label{flowtri2x1cyc}
\eeq
\beq
F(tri[2 \times 2,cyc.],q) = (q-1)(q^4-7q^3+21q^2-30q+17)
\label{flowtri2x2cyc}
\eeq
\beq
F(tri[2 \times 3,cyc.],q) = (q-1)(q^6-11q^5+55q^4-159q^3+282q^2-290q+133)
\label{flowtri2x3cyc}
\eeq
\beqs
F(tri[2 \times 4,cyc.],q) & = & (q-1)(q^8-15q^7+105q^6-447q^5+1269q^4
\cr\cr
& & -2463q^3+3185q^2-2494q+897)
\label{flowtri2x4cyc}
\eeqs

\subsection{$L_{\lowercase{y}}=3$ Cyclic Strips of the Triangular Lattice}

\beq
F(tri[3 \times 1,cyc.],q) = (q-1)^5
\label{flowtri3x1cyc}
\eeq
\beq
F(tri[3 \times 2,cyc.],q) =
(q-1)(q^2-3q+3)(q^6-10q^5+45q^4-115q^3+177q^2-157q
+63)
\label{flowtri3x2cyc}
\eeq
\beqs
& & F(tri[3 \times 3,cyc.],q) = (q-1)(q^{12}-20q^{11}+190q^{10}-1134q^9
\cr\cr
& & +4743q^8-14683q^7+34624q^6-62969q^5+88184q^4-93366q^3+71377q^2
\cr\cr
& & -35497q+8671)
\label{flowtri3x3cyc}
\eeqs

\vfill
\eject
\end{document}